\documentclass{aa}  
\usepackage{graphicx}
\usepackage{txfonts}
\usepackage{xcolor}
\usepackage{subfig}
\usepackage{makecell}
\usepackage{color}
\usepackage{here}
\usepackage{comment}
\usepackage{cleveref}
\begin{document}

\title{Photometry and high-resolution spectroscopy of comet 21P/Giacobini-Zinner during its 2018 apparition
   \thanks{Based on observations collected at the European Southern Observatory under ESO program 2101.C-5051.}}

\authorrunning{Moulane et al. 2020}
\titlerunning{Comet 21P/Giacobini-Zinner}
        
\author{
Y. Moulane\inst{1,2,3}$^\dagger$, 
E. Jehin\inst{2}, 
P. Rousselot\inst{4}, 
J. Manfroid\inst{2}, 
Y. Shinnaka\inst{5},
F. J. Pozuelos\inst{2},
D. Hutsemékers\inst{2}, 
C. Opitom\inst{1,6}, 
B. Yang\inst{1}, \and
Z. Benkhaldoun\inst{3} 
           }
           
\institute{
European Southern Observatory, Alonso de Cordova 3107, Vitacura, Santiago, Chile\\
$^\dagger$\email{\color{blue}ymoulane@eso.org}\and
Space sciences, Technologies \& Astrophysics Research (STAR) Institute, University of Liège, Liège, Belgium\and
Oukaimeden Observatory, High Energy Physics and Astrophysics Laboratory, Cadi Ayyad University, Marrakech, Morocco\and
Institut UTINAM UMR 6213, CNRS, Univ. Bourgogne Franche-Comté, OSU THETA, BP 1615, 25010 Besançon Cedex, France\and
Koyama Astronomical Observatory, Kyoto Sangyo University, Motoyama, Kamigamo, Kita-ku, Kyoto 603-8555, Japan \and
Institute for Astronomy, University of Edinburgh, Royal Observatory, Edinburgh EH9 3HJ, UK\\
             }

\date{Received/accepted}

\abstract{We report on photometry and high resolution spectroscopy of the chemically peculiar Jupiter-family Comet (hereafter JFC) 21P/Giacobini-Zinner. Comet 21P is a well known member of the carbon-chain depleted family but displays also a depletion of amines. We monitored continuously the comet over more than seven months with the two TRAPPIST telescopes (TN and TS), covering a large heliocentric distance range from 1.60 au inbound to 2.10 au outbound with a perihelion at 1.01 au on September 10, 2018. We computed and followed the evolution of the dust (represented by Af$\rho$) and gas production rates of the daughter species OH, NH, CN, C$_3$, and C$_2$ and their relative abundances to OH and to CN over the comet orbit. We compared them to those measured in the previous apparitions. The activity of the comet and its water production rate reached a maximum of (3.72$\pm$0.07)$\times$10$^{28}$ molec/s on August 17, 2018 (r$_h$=1.07 au), 24 days before perihelion. The peak value of A(0)f$\rho$ was reached on the same date (1646$\pm$13) cm in the red filter. Using a sublimation model for the nucleus, we constrained the active surface of the nucleus using the slow-rotator model. The abundance ratios of the various species are remarkably constant over a large range of heliocentric distances, before and after perihelion, showing a high level of homogeneity of the ices in the surface of the nucleus. The behaviour and level of the activity of the comet is also remarkably similar over the last five orbits. About the coma dust colour, 21P shows reflectively gradients similar to JFCs. We obtained a high resolution spectrum of 21P with UVES at ESO VLT one week after perihelion. Using the CN B-X (0, 0) violet band, we measured $^{12}$C/$^{13}$C and $^{14}$N/$^{15}$N isotopic ratios of 100$\pm$10 and 145$\pm$10, respectively, both in very good agreement with what is usually found in comets. We measured an ortho-para abundance ratio of NH$_3$ of 1.16$\pm$0.02 corresponding to a nuclear spin temperature of T$_{spin}$=27$\pm$1 K, similar to other comets. While the abundance ratios in the gaseous coma reveal a peculiar composition, the isotopic and ortho-to-para ratios appear totally normal.  We performed a dynamical simulation of 21P and we found that it is likely a young member of the JFC population. We favour a pristine composition scenario to explain the chemical peculiarities of this comet.}


\maketitle
\section{Introduction}
\label{intro}

Comet 21P/Giacobini-Zinner (hereafter 21P) is a JF with a short period of 6.5 years. 21P was discovered in 1900 by Michel Giacobini and rediscovered by Ernst Zinner in 1913\footnote{\url{https://ssd.jpl.nasa.gov/sbdb.cgi?sstr=21P;old=0;orb=0;cov=0;log=0;cad=0\#discovery}}. After its discovery, 21P was observed in most of its apparitions and many photometric and spectroscopic measurements were reported. In September 1985, 21P was the first comet visited by  the International Cometary Exporer (ICE) spacecraft to study the interaction between the solar wind and the cometary atmosphere \citep{Von1986,Scarf1986}. 21P is also known as the parent body of the Draconids meteor shower \citep{Beech1986,Egal2019}. Many spectroscopic-photometric studies at various wavelength ranges have been performed since its discovery \citep{Schleicher1987,Cochran1987,Fink1996,Weaver1999,Lara2003,Combi2011} including production rates measurements, atomic and molecular abundances. 21P is the prototype of depleted comets in C$_2$ and C$_3$ with respect to CN and to OH \citep{Schleicher1987,AHearn1995}.  C$_2$ and C$_3$ relative abundances are about 5 and 10 times smaller than those measured in “typical” comets \citep{AHearn1995}. 21P was found to be also depleted in NH \citep{Schleicher1987,Kiselev2000} and NH$_2$ \citep{Konno1989,Beaver1990,Fink2009}.

The physical properties of 21P were investigated  in the previous apparitions. Its nucleus size is not well determined, the average estimate of its radius is about 1-2 km \citep{Tancredi2000,Krolikowska2001,Pittichova2008}. 
Its rotation period is not well constrained, a large range from 9.5 hr to 19 hr was estimated \citep{Leibowitz1986}. 21P activity has shown an asymmetric light curve with respect to perihelion in the previous apparitions. In 1985, the production rates pre-perihelion were two times larger than post-perihelion at heliocentric distances of 1.0-1.5 au \citep{Schleicher1987}. The gas and dust maximum production was observed about one month before perihelion for the previous apparitions \citep{Schleicher1987,Hanner1992,Lara2003}. Both its unusual composition and the behavior of its activity during multiple apparitions make 21P an object of great interest. In addition, as it is the parent body of the Draconids, a study of its dust properties might be valuable. The 2018 apparition was very favorable for ground-based observations since the comet was close both to the Sun and the Earth at the same time and reaching high elevation.  Many observed the comet again for the 2018 return using various state of the art  instrumentation (IR and optical spectrographs on large telescopes) in order to better  understand  these peculiarities. 

This work is organized as follows: after the introduction and historical background given in section \ref{intro}, we describe the observing circumstances and the reduction process of TRAPPIST images and UVES/VLT spectra in section \ref{data_reduction}. In section \ref{activity}, we compute the production rates and we discuss the gas and dust activity pre- and post-perihelion as well as the properties of the dust.  The relative molecular abundances and their evolution with respect to the heliocentric distance are discussed in section \ref{abundances} and compared to the IR mother species abundances. In section \ref{sec_ratios}, we present the nitrogen and carbon isotopic ratios and the NH$_2$ (and NH$_3$) ortho-para ratio derived from the UVES high resolution spectrum.  In section \ref{sec_dynamical}, we investigate the dynamical evolution of comet 21P within the last 10$^5$ years. Discussion of the chemical composition of 21P and the possible scenarios of its depletion in carbon species are given in section \ref{discussion}. The summary and conclusions of this work are given in the last section \ref{sec_conclusion}.

\section{Observation and data reduction}
\label{data_reduction}
 
\subsection{Photometry (TRAPPIST)}
\label{Photometry_TRAPPIST}
We started monitoring 21P with TRAPPIST-North (hereafter TN) at the beginning of June 2018 when the comet was at 1.55 au from the Sun. The comet was then observed from the southern hemisphere with TRAPPIST-South (hereafter TS) from the beginning of September 2018. The pair of TRAPPIST telescopes \citep{Jehin2011} is in such a case very useful as it allowed a continuous monitoring of the comet before and after perihelion. We collected images with the cometary HB narrow-band filters \citep{Farnham2000} to measure the production rates of the radicals OH, NH, CN, C$_3$, and C$_2$. We also acquired images with the dust continuum filters BC, GC, and RC for blue, green and red \citep{Farnham2000}. We used the broad-band filters B, V, Rc, and Ic \citep{Bessell1990} to compute the Af$\rho$ parameter  which is a proxy of the dust production rate \citep{A'Hearn1984} and to derive the  dust colours. 

Throughout the passage of the comet, we made a high cadence monitoring of 21P with images taken about twice a week. On photometric nights, we also obtained long series of observations with the gas narrow-band filters, especially CN and C$_2$ filters, to measure the variations of the production rates during the same night due to the rotation of the nucleus. We chose the exposure time of the different filters depending on the brightness of the comet. We used exposure times between 60 s and 240 s for the broad-band filters and between 600 s and 1500 s for the narrow-band filters. Observational circumstances and number of sets of each filter are summarized in Table \ref{circ} in the appendix. We started to collect data three months before perihelion to four months after perihelion. The comet reached perihelion on September 10, 2018 at a heliocentric distance of 1.01 au and a geocentric distance of 0.39 au. In total the comet was observed on 50 different nights with both telescopes, 13 nights before perihelion and 37 after.
We used the same procedures described in our previous papers (e.g, \cite{Opitom2015a,Moulane2018} and references therein) to reduce the data and to perform the flux calibration. 
To compute the production rates, we converted the flux of different gas species to column densities and we adjusted their profiles with a Haser model \citep{Haser1957}.  This simple model, but widely used, is based on a number of assumptions. Outgassing is assumed to be isotropic and the gas has a constant radial velocity of 1 km/s. Parent molecules coming off the nucleus are decaying by photo dissociation to produce the observed daughter molecules. The model adjustment was performed at a physical distance of $10\,000$~km from the nucleus. 
Table \ref{tab:parameters} shows the scale lengths and g-factors of different molecules at 1 au scaled by r$_h^{-2}$. More details about the Haser model and its parameters are given in our previous works  (see \cite{Moulane2018}  and references therein). We would like to point out that we used the same parameters as those from \cite{Schleicher2018} for the previous apparitions of 21P.
We derived the Af$\rho$ parameter, a proxy for the dust production \citep{A'Hearn1984}, from the dust profiles in the cometary dust continuum BC, GC, and RC filters and the broad-band Rc and Ic filters. It was computed at $10\,000$~km from the nucleus and corrected for the phase angle effect according to the phase function normalized at $\theta$=0$^\circ$ derived by D.~Schleicher\footnote{\url{http://asteroid.lowell.edu/comet/dustphase.html}}.

\begin{table}
        \begin{center}
                \caption{The scale lengths and the fluorescence efficiency of different molecules at 1 au scaled by r$_h^{-2}$.}
                
                \begin{tabular}{lccc} 
                        \hline
                        Molecules  &  Parent   & Daughter    & g-factors \\
                                   &  (km)     &  (km)       & erg s$^{-1}$ mol$^{-1}$\\
                        \hline 
                        OH(0,0) &  2.4 $\times$10$^4$ & 1.6 $\times$10$^5$ & 1.49$\times$10$^{-15}$\\
                        NH(0,0) &  5.0 $\times$10$^4$ & 1.5 $\times$10$^5$ & 6.27$\times$10$^{-14}$\\
                        CN($\Delta\upsilon$=0) &  1.3 $\times$10$^4$ & 2.1 $\times$10$^5$ & 2.62$\times$10$^{-13}$ \\
                        C$_3$($\lambda$=4050 \AA{}) &  2.8 $\times$10$^3$ & 2.7 $\times$10$^5$ & 1.00$\times$10$^{-12}$ \\
                        C$_2$($\Delta\upsilon$=0) & 2.2 $\times$10$^4$ & 6.6 $\times$10$^4$&4.50$\times$10$^{-13}$ \\
                        \hline  
                        \hline
                        \label{tab:parameters} 
                \end{tabular}
                \vspace{-0.5cm}
                \tablefoot{The scale lengths are equivalent to the lifetimes of molecules as we are using a constant radial velocity of 1 km/s \citep{AHearn1995}. The fluorescence efficiency are taken from Schleicher's website \footnote{\url{https://asteroid.lowell.edu/comet/gfactor.html}}. }
        \end{center}
\end{table}

\subsection{Spectroscopy (UVES/VLT)}

We obtained one spectrum of comet 21P with the Ultraviolet-Visual Echelle Spectrograph (UVES) mounted on the Unit 2 telescope (UT2) at ESO's VLT on September 18, 2018 (a week after perihelion, r$_h$=1.01 au and $\Delta$=0.40 au) under Director’s Discretionary Time.  We used the UVES standard settings DIC\#1 346+580 covering the range 3030 to 3880 \AA~in the blue and 4760 to 6840 \AA~ in the red.
We used a 0.44$^{\prime\prime}$ wide slit, providing a resolving power R$\sim$80 000. 
We obtained one single exposure of 3000 s at 8h35 UT with a mean airmass of 1.7. This exposure provided two different spectra, both of them covering one of the above mentioned spectral ranges.

\begin{figure*}
\begin{center}
	\includegraphics[scale=0.65,angle=-90]{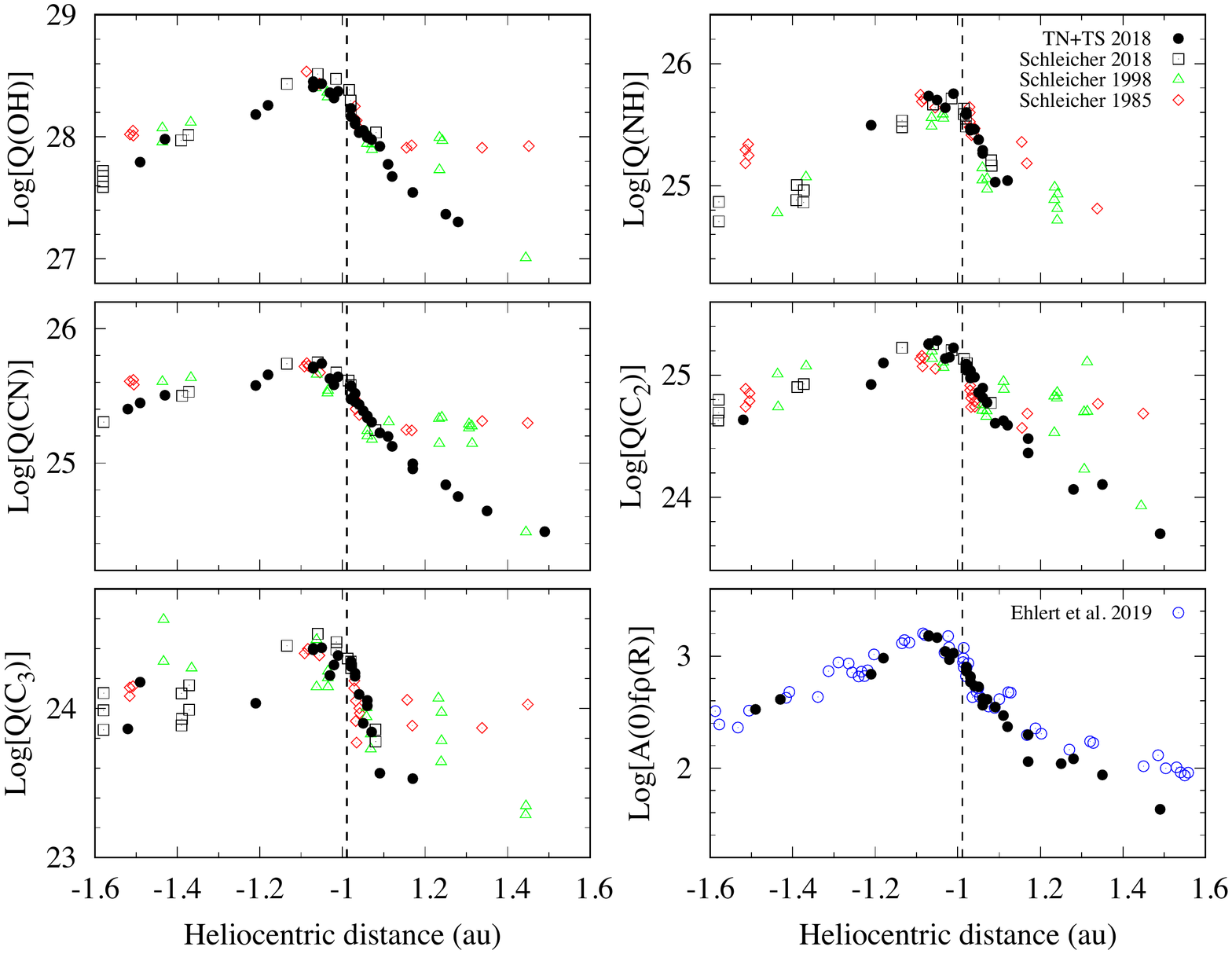}
	\caption{The logarithm of the production rates (in molec/s) of each observed species and of the A(0)f$\rho$ parameter (in cm), of comet 21P during its 2018 return (this work and \cite{Schleicher2018}) are compared with two previous apparitions in 1985 and 1998 \citep{Schleicher2018} as a function of heliocentric distance. The dashed vertical line represents the perihelion distance on September 10, 2018. The maximum of the gas and dust activity was reached at 1.07 au from the Sun, on August 17, 2018, 24 days before perihelion. }
	\label{fig:gas_rate}
\end{center}
\end{figure*}

The ESO UVES pipeline was used to reduce the spectra in the extended object mode, keeping the spatial information. The spectra were corrected for the extinction and flux calibrated using the UVES master response curve provided by ESO. One-dimensional spectra were then extracted by averaging the 2D spectra with simultaneous cosmic ray rejection and then corrected for the Doppler shift due to the velocity of the comet with respect to the Earth. More details about the UVES data reduction are given in the UVES manual\footnote{\url{ftp://ftp.eso.org/pub/dfs/pipelines/uves/uves-pipeline-manual-22.17.pdf}}. The dust-reflected sunlight was finally removed using a reference solar spectrum BASS2000\footnote{\url{http://bass2000.obspm.fr/solar_spect.php}}. More in depth description of the steps for UVES data reduction and the solar spectrum subtraction is given in \cite{Manfroid2009} and references therein.

\section{Activity and composition}
\label{activity}

\subsection{Gas production rates}

Along with our monitoring of comet 21P with the TRAPPIST telescopes, we derived the OH, NH, CN, C$_3$, and C$_2$ production rates. They are summarized in Table \ref{tab:rates} and their evolution as a function of the heliocentric distance are compared with two previous passages in Figure \ref{fig:gas_rate}.

We started to detect most of the radicals in the coma by the end of June 2018 (except NH one month later). The various production rates and the dust activity have been increasing slowly as the comet was getting closer to the Sun (from 1.52 to 1.07 au). The maximum of the activity was reached at 1.07 au from the Sun, on August 17, 24 days before perihelion. It then started to decrease rapidly after perihelion. CN was detected in our data until the end of 2018 at 1.66 au, while OH, C$_3$, and C$_2$ were not detected anymore after the beginning of November at 1.4 au and NH early October at 1.2 au. We found that like for the previous apparitions \citep{Schleicher1987}, the production rates pre-perihelion are larger by more than a factor two than post-perihelion. It is clear from Figure \ref{fig:gas_rate} that the asymmetric activity is seen for all species and this behaviour does not change over the various apparitions \citep{Schleicher1987,Combi2011} as shown also in Figure \ref{fig:rate_H2O} for the water production rate. The same behavior has been reported for the parent molecules (H$_2$O, CO, CH$_4$, C$_2$H$_2$, C$_2$H$_6$, NH$_3$ and CH$_3$OH) production rates derived at infrared wavelengths during the 2018 passage \citep{Faggi2019,Roth2020}. This might 
due to the shape of the nucleus and its spin-axis orientation. This effect has been observed in several comets such as 9P/Tempel 1 \citep{Schleicher2007}, 81P/Wild 2 \citep{Farnham2005} and also for comet 67P/Churyumov-Gerasimenko \citep{Schleicher2006,Opitom2017}. It has been shown very clearly by Rosetta mission that the maximum activity of 67P was well associated to the illumination of the most southern regions, which were receiving the maximum of solar flux, and were subject to intense erosion \citep{Lai2019}. 
Recently, \cite{Marshall2019} has show for those three comets that the nucleus shape, the spin axis orientation, and the distribution of activity on the comet’s surface can explain the water production rate light-curve as a function of the heliocentric distance.

Around the maximum of the activity, the production rates are almost the same as those measured in the previous apparitions showing that there is no decrease of the activity level of 21P over the last five orbits. Our production rates agree usually very well with those derived by \cite{Schleicher2018} who used the same technique, while we noticed a discrepancy at large heliocentric distance and post-perihelion in the 1985 and 1998 apparitions data \citep{Schleicher1987,Schleicher2018}. This could be due to a sensitivity issue in their data as the production rates seem to level off on both sides of perihelion while the distance increase or to a higher activity level of the comet after perihelion in the past. It has been also found that there is no significant change in the production rates of hypervolatile molecules (CO, CH$_4$, and C$_2$H$_6$) in comet 21P over the three different apparitions, 1998 \citep{Weaver1999,Mumma2000}, 2005 \citep{DiSanti2012} and 2018 \citep{Faggi2019,Roth2020}.

\subsection{ H$_2$O production rate}
\label{water-production-rate}
\begin{figure*}[h!]
\centering	\includegraphics[scale=0.65,angle=-90]{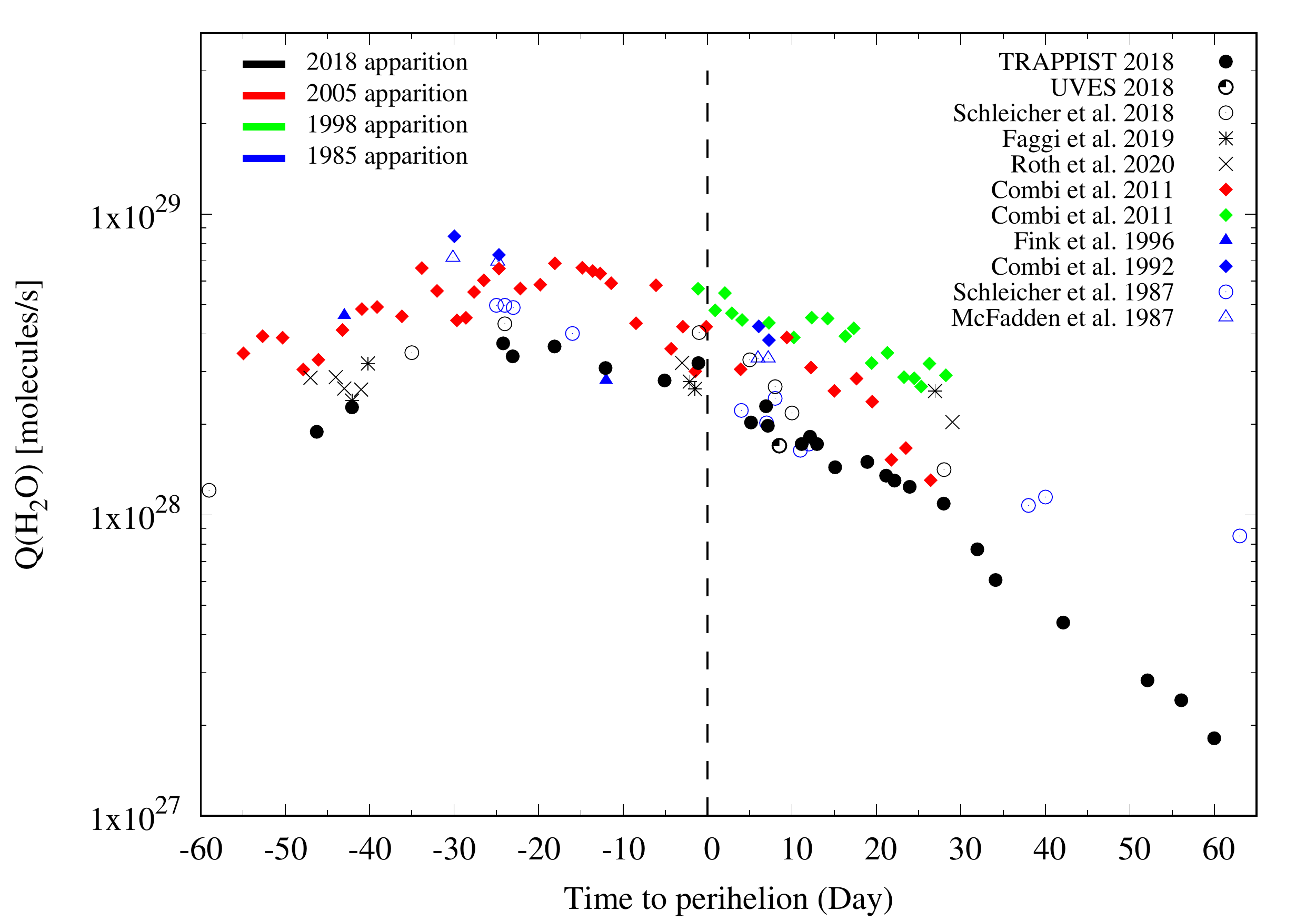}
	\caption{H$_2$O production rates of comet 21P as a function of days to perihelion in 2018 compared to the previous apparitions (1985, 1998 and 2005). More description about the data is given in section \ref{water-production-rate}. }
	\label{fig:rate_H2O}
\end{figure*}

The water production rate is the most significant indicator of the activity of a comet. It can be measured directly from near-infrared observations or derived from OH emission at 3090 \AA~ and radio wavelengths or from H Lyman-$\alpha$ emission at 1216~$\AA$ \citep{Combi1986} assuming that both OH and H arise from the dissociation of H$_2$O. In this work, we computed the vectorial-equivalent water production rates according to an empirical procedure based on a comparison of OH and water production rates derived from the mean lifetimes, velocities, and scale lengths given by \cite{Cochran1993}. \cite {Schleicher1998} built an empirical relationship $Q(H_2O)=1.361 r_h^{-0.5} Q(OH)$ based on a r$_h^{-0.5}$ dependence of the H$_2$O outflow velocity, a photodissociation branching ratio for water to OH of 90\%, and the heliocentric distance. Figure \ref{fig:rate_H2O} shows the water production we derived compared to previous apparitions (with different techniques) as a function of days to perihelion. We used the formula given above to convert Q(OH) to water production rates for \cite{Schleicher2018} data. \cite{Combi2011} derived the production rates from the H Ly-$\alpha$ emission observed by the SWAN instrument on board SOHO in 1998 and in 2005. \cite{Combi1992} values are derived from H Ly-$\alpha$  emission observed by the IUE mission for the 1985 apparition. From the Pioneer Venus Orbiter ultraviolet system (UVS) instrument, \cite{McFadden1987} derived the water production rates from OH (3090 \AA) emission. \cite{Fink1996} derived the water production rates from the [OI]($^1$D) forbidden line doublet using a correlation between the water production rates and the total photon luminosity. \cite{Faggi2019} and \cite{Roth2020} measured directly the water production rates from near-infrared spectra. TRAPPIST and UVES data points are from this work (see section \ref{water-production-rate} and \ref{uves-water-production-rate}). The maximum in the last four apparitions was reached about one month before perihelion and does not change over all apparitions, but we observe a clear systematic difference between the narrow band and spectroscopic methods in the optical on one hand and the measurements made from the space observations of the H Ly-$\alpha$ emission in the UV on the other hand.  The maximum of the water production we measured was on August 17, 24 days before the perihelion, and it reached (3.72$\pm$0.07)$\times$10$^{28}$ molec/s in good agreement with \cite{Schleicher2018} measurement of 4.20$\times$10$^{28}$ molec/s at the heliocentric distance 1.07 au. Using the same technique for the 1985 apparition, \cite{Schleicher1987} reported Q(H$_2$O)=4.85$\times$10$^{28}$ molec/s when the comet was at 1.05 au from the Sun. Using high resolution infrared spectroscopy, \cite{Weaver1999} measured $\sim$2–3$\times$10$^{28}$ molec/s at r$_h$=1.10 au in 1998. For the 2005 apparition, \cite{Combi2011} found a value of 5.80$\times$10$^{28}$ molec/s from the H Ly-$\alpha$ emission observed by the SWAN/SOHO at r$_h$=1.08 au. Comparing these data, we found that the water production rates measured by \cite{Combi1992} in 1985 and \cite{Combi2011} in 2005 are systematically higher by a factor of about two with respect to our results in 2018. Such an offset between various techniques has been reported in previous studies and as early as comet 1P/Halley \citep{Schleicher1998}. The origin of this discrepancy is not clear but it is obvious that there is a good agreement when the same technique is used. This indicates that the level of activity of 21P was the same over the past four decades and did not decrease like comet 41P/Tuttle–Giacobini–Kresak which has been losing as much as 30\% to 40\% of its activity from one orbit to the next \citep{Moulane2018}.

\subsection{Active area of the nucleus}
\label{active-area}

To estimate the active area of the nucleus' surface, we modelled the water production using the sublimation model of \cite{cowan1979}. Due to the low thermal inertia of cometary nuclei \citep{gulkis2015}, the slow-rotator approach was adopted in a number of cases as the most appropriate way to compute the cometary outgassing \cite[see e.g.,][]{bodewits2014,darius2019}. The slow-rotator model assumes every facet of the nucleus is in equilibrium with the solar radiation incident upon it, with the rotational pole pointed at the Sun. As mentioned previously, the size of 21P's nucleus, necessary to convert the active area to the active fraction of the whole surface, is not well constrained so far, with a radius ranging from 1 to 2~km. Hence, to estimate the active fraction of the surface we assumed a radius of 1.5$\pm$0.5~km. Moreover, we assumed a bond albedo of 5$\%$ and a 100$\%$ infrared emissivity \cite[see e.g.,][]{hearn1989,mckay2018,mckay2019}. We found that the active area of 21P during our monitoring campaign varied from $\sim$5~km$^{2}$ at 1.49~au pre-perihelion, reached a maximum of $\sim$12~km$^{2}$ at 1.07~au pre-perihelion, and decreased to $\sim$1~km$^{2}$ at 1.31~au post-perihelion. Table~\ref{tab:subli} shows the minimum and maximum active areas and active fraction for 21P using the slow-rotator model at some interesting heliocentric distances. We obtained different values in comparison with previous estimations given by \citet{combi2019}. The reason is twofold; first, the already mentioned discrepancy in the water production rates found via different observational techniques (see Section 3.2), and secondly, the model used by the authors (fast-rotator), which is less appropriate to describe the cometary outgassing.

\begin{table}[h!]
	\caption{ Active area (km$^{2}$) and active fraction of the surface ($\%$) for 21P using the slow-rotator model at some interesting heliocentric distances.}
	\label{tab:subli}
	\resizebox{0.49\textwidth}{!}{%
		\begin{tabular}{lccc}
			\hline
			\hline
			\rule{0pt}{2ex} Date UT & r$_h$  & Active Area  & Active fraction   \\ 
			        &  (au) & (km$^{2}$)  & ($\%$)\\
		    \hline
		    \rule{0pt}{2ex}(a) 2018 Jun 22       &-1.49 & 4.9$\pm$0.1     & 17.5$\pm$11.6 \\
			(b) 2018 Aug 17       & -1.07 & 12.0$\pm$0.2   & 42.5$\pm$28.0 \\
			(c) 2018 Sep 09       & -1.01 & 9.0$\pm$0.2    & 32.0$\pm$21.3  \\
			(d) 2018 Sep 15,16,17 & +1.02 & 6.9$\pm$0.1    & 21.5$\pm$14.0    \\
			(e) 2018 Nov 09       & +1.31 & 0.90$\pm$0.04  &  3.3$\pm$2.2    \\
			\hline
			\hline
	\end{tabular}}
	\tablefoot{ (a) The first measurement during our monitoring campaign, (b) the maximum activity during our monitoring campaign, (c) the last measurement before perihelion passage, (d) the mean of the first measurements after perihelion passage with similar heliocentric distances of 1.02~au, and (e) the last and minimum measurements during our monitoring campaign. Large errors in the active fractions of the surface came from the large uncertainties in the radius of the nucleus, which we adopted as 1.5$\pm$0.5~km (see section \ref{active-area} for details).}
\end{table}

\subsection{Dust properties}
\label{sec_colors}

\begin{figure}[h!]
	\centering	\includegraphics[width=0.72\columnwidth,angle=-90]{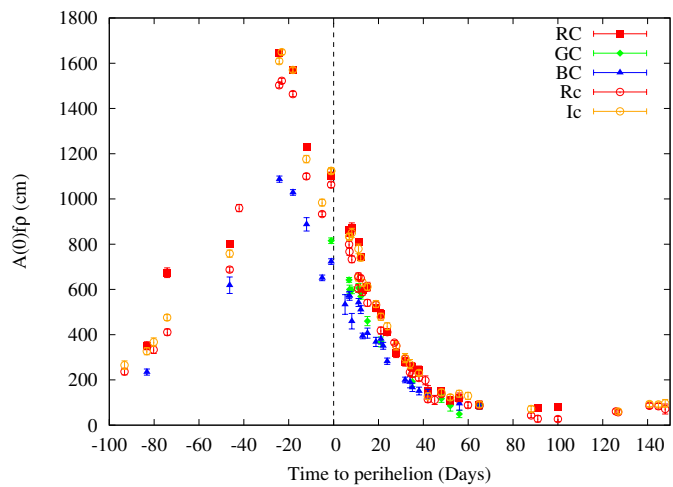}
	\caption{The A(0)f$\rho$ parameter measurements, computed at 10~000~km from the nucleus and corrected for the phase angle effect, for the broad band (Rc and Ic) and narrow-band cometary filters (RC,GC,BC) as a function of days to perihelion.} 
	\label{fig:afrho_BVRI}
\end{figure}

We computed the A(0)f$\rho$ parameter at 10000\,km, as defined by \cite{A'Hearn1984}, using broad-band (Rc and Ic) and narrow-band dust continuum filters (RC,GC,BC) (see Table \ref{tab:rates}). Figure \ref{fig:afrho_BVRI} shows its evolution as a function of time to perihelion. Our results are in very good agreement with those reported by \cite{Ehlert2019}, as shown in the right bottom panel of Figure \ref{fig:gas_rate}. Like for the gas, the maximum was reached on August 17 with a value of (1646.1$\pm$12.8) cm in the red narrow-band RC filter. About the same value was reported on previous apparitions at the same heliocentric distance \citep{Schleicher1987,Lara2003,Pittichova2008}. For a detailed description of the dust environment and its evolution, a more sophisticated model should be used, such as the Monte Carlo model presented by \cite{moreno2012} which was used successfully in a number of cases \citep[see e. g., ][]{pozuelos2015,moreno2016a,moreno2016b,moreno2017,pozuelos2018}. Such a study will be presented in a separate paper. 

We used the A(0)f$\rho$ values obtained with the narrow-band filters which are not contaminated by the gas emissions to derive the dust colours. The normalized reflectivity gradients between wavelength $\lambda_1$ and $\lambda_2$ is defined as \citep{A'Hearn1984,Jewitt1986}:

\begin{equation}
S_v(\% /1000 \AA)=\frac{Af\rho_1-Af\rho_2}{Af\rho_1+Af\rho_2}\times\frac{2000}{\lambda_1-\lambda_2}
\end{equation}

$\lambda_1$ and $\lambda_2$ are the effective wavelengths of the filters: BC[4450 \AA], GC[5260 \AA] and RC[7128 \AA]. 

\begin{figure}[h!]
	\includegraphics[width=0.72\columnwidth,angle=-90]{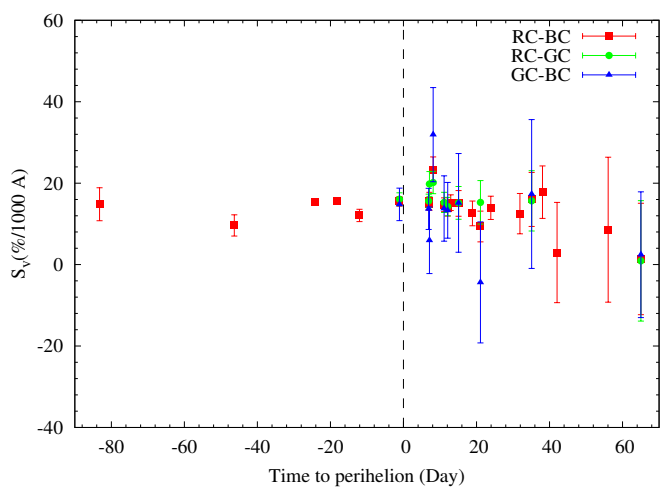}
	\caption{Normalized reflectivity gradients $S_v$(\% per 1000 \AA) of comet 21P for different colour indices as a function of days to perihelion. } 
	\label{color_slope}
\end{figure}

We found that the RC-GC, RC-BC and GC-BC colours are redder than the Sun with mean values of (14.8$\pm$3.3), (13.2$\pm$2.6) and (12.4$\pm$7.5)\%/1000 $\AA$ respectively (see Figure \ref{color_slope}). This result is in agreement with previous apparitions, with values of S$_v$=15\%/1000 $\AA$ in 1985 \citep{Schleicher1987} and S$_v$=13\%/1000 $\AA$ in 1998 \citep{Lara2003}. These values are consistent with the colour of the nucleus of 21P (12.8$\pm$2.7)\%/1000 $\AA$ measured at a heliocentric distance of 3.5 au in 1991 \citep{Luu1993}. They fall within the range observed for most JFCs  \citep{Lamy2009,Solontoi2012,Jewitt2015}. 
During our long monitoring, we did not detect any significant variation of the colour of the dust in the coma (or any outburst). 

\section{Abundances ratios}
\label{abundances}

\begin{figure*}[h!]
	 \includegraphics[scale=0.65,angle=-90]{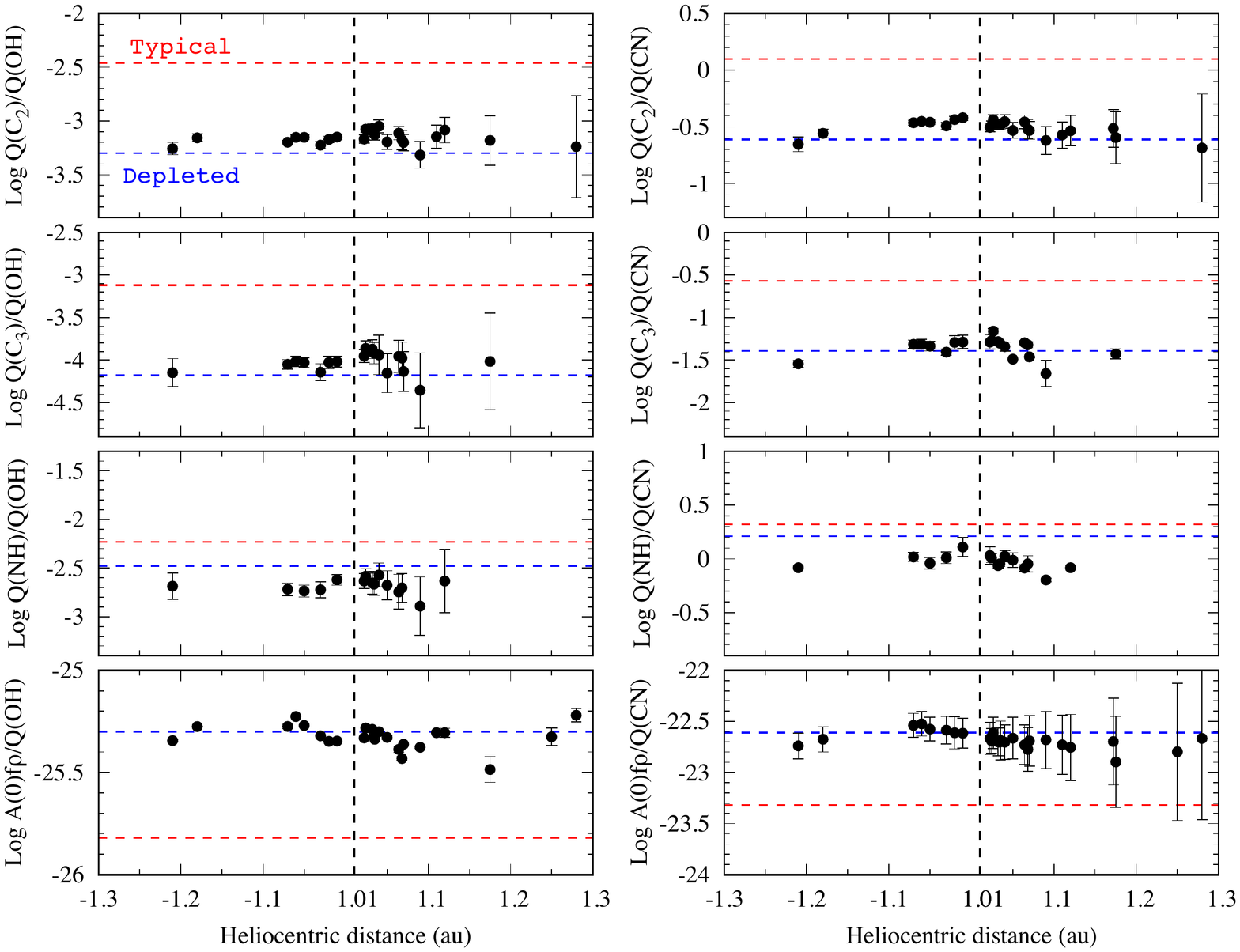}
		\vspace{-0.5cm}
	\caption{Evolution of the logarithmic production rates ratios of each species with respect to OH and to CN as a function of heliocentric distance. The red dashed line represents the mean value of typical comets as defined in \cite{AHearn1995} while the blue one represents the mean value of the depleted group. The bottom panels shows the dust-to-gas ratio represented by A(0)f$\rho$-to-OH and A(0)f$\rho$-to-CN. The vertical dashed line shows the perihelion distance on September 10, 2018 at r$_h$=1.01 au.} 
	\label{ratios}
\end{figure*}

Studying the molecular abundances and their ratios with respect to the distance to the Sun gives information about the homogeneity of a comet's nucleus and the chemical processes involved in the coma. 
Based on the relative abundance of 41 comets, \cite{AHearn1995} classified comets into two groups based on their C$_2$/CN ratio. Typical comets are defined as those having a Log[Q(C$_2$)/Q(CN)] $\geq$ -0.18 while the carbon-chain depleted comets are those below that value. This classification was confirmed later by other photometric and spectroscopic studies of large data sets \citep{Schleicher2008,Fink2009,LanglandShula2011,Cochran2012} and must reflect some differences between the formation conditions (the pristine scenario) or a change of  relative composition with time (several perihelion passages) of these comets (the evolutionary scenario). 
Figure \ref{ratios} shows the evolution of the 21P abundance ratios of the various radicals with respect to OH (a proxy of water) and CN, as a function of heliocentric distance. It is clear that 21P abundance ratios in the 2018 return agree with the mean values of depleted comets given in \cite{AHearn1995}. Table \ref{tab:abundance} summarizes the relative abundances in 2018 compared to 1985 and 1998 data using the same technique and the same Haser model parameters \citep{Schleicher2018}. Our 2018 ratios are the mean values for all the data obtained (see Table \ref{tab:rates}). Like for the activity level over the past passages, the relative abundances did not change over the last 5 orbits. We note that the Af$\rho$ values derived in 1985 and in 1998 by \cite{Schleicher2018} were computed for the narrow band GC[5260 \AA] filter while we used the BC[4450 \AA] filter. After correcting their Af$\rho$ values for the phase angle effect using the same function as for the TRAPPIST data (see section \ref{Photometry_TRAPPIST}), both data sets are in agreement. This indicates that there is no evidence of changes in the chemical composition in the coma of the comet at different heliocentric distances (in the range 1.0 to 1.5 au) and over the five orbits, which is an argument to reject the evolutionary origin of the carbon chain depletion in that comet.

\begin{table}[h!]
	\begin{center}
		\caption{Relative molecular abundances of comet 21P over the last passages compared to the mean values for typical comets.}
		\label{tab:abundance}
		\resizebox{0.49\textwidth}{!}{%
			\begin{tabular}{lcccc}
					\hline	
				\hline
			 \multicolumn{5}{c}{Log production rate ratio}\\
				\hline
				& 1985$^{(a)}$ & 1998$^{(a)}$& 2018$^{(b)}$  & Typical comets$^{(c)}$\\
				 \hline
			  C$_2$/CN &   -0.64 &-0.50 &  -0.52$\pm$0.10   &  0.06$\pm$0.10 \\
			  C$_3$/CN &   -1.42 &-1.30 &  -1.39$\pm$0.12   & -1.09$\pm$0.11 \\
			  CN/OH    &   -2.59 &-2.67 &  -2.62$\pm$0.08   & -2.50$\pm$0.18 \\
              C$_2$/OH &   -3.23 &-3.17 &  -3.16$\pm$0.21   & -2.44$\pm$0.20 \\
              C$_3$/OH &   -4.02 &-3.98 &  -4.03$\pm$0.16   & -3.59$\pm$0.29 \\
             NH/OH      &  -2.66 &-2.87 &  -2.68$\pm$0.14   & -2.37$\pm$0.27 \\
             A(0)f$\rho$/CN & -22.74&-22.73 &  -22.70$\pm$0.04  & -23.30$\pm$0.32\\
             A(0)f$\rho$/OH & -25.33&-25.42 &  -25.32$\pm$0.04  & -25.82$\pm$0.40\\
				\hline	
				\hline
			\end{tabular}}
		\end{center}
		\vspace{-0.5cm}
	\tablefoot{$^{(a)}$\cite{Schleicher2018}, $^{(b)}$ This work, $^{(c)}$ \cite{AHearn1995}. }
	\end{table}
	
As seen in the bottom panel of Figure \ref{ratios}, there is also no evidence that the dust-to-gas ratio represented by A(0)f$\rho$/Q(CN) and A(0)f$\rho$/Q(OH) depends on the heliocentric distance. We found that this ratio in 21P is consistent with the average value of depleted comets and higher than the mean value of the typical comets as defined in \cite{AHearn1995} (see Table \ref{tab:abundance}). \cite{Lara2003} obtained a value of Log[A(0)f$\rho$/Q(CN)]=-22.91$\pm$0.10 in 1998 which is in agreement with our measurement. Like for the gas relative abundances, we conclude that 21P coma does not show significant variation in the dust-to-gas ratio over the previous apparitions and it has a similar ratio as the depleted comets defined by \cite{AHearn1995}.

\begin{table*}
	\caption{Comparison of daughter molecules and possible parent molecules production rates derived from optical and infrared data of comet 21P in the 2018 passage.}
	\label{tab:infrared-optical}
	\resizebox{\textwidth}{!}{%
		\begin{tabular}{lcccccccccccl}
			\hline
			\hline
			UT Date   & $r_h$ & $\bigtriangleup$  &  \multicolumn{9}{c}{Production rates  (10$^{25}$molec/s)}    & Reference \\
			&(au) & (au)  & Q(OH) &Q(H$_2$O) & Q(CN) & Q(HCN)	& Q(C$_2$) & Q(C$_2$H$_2$) &Q(C$_2$H$_6$) & Q(NH) & Q(NH$_3$) & \\
			\hline
		
			2018 Jul 30 &1.17&0.61 &1810$\pm$28 & &4.55$\pm$0.07 & -&1.26$\pm$0.07 & -&- &3.13$\pm$0.50 &- & This Work\\
			2018 Jul 30 &1.17&0.61 &- &2401$\pm$394 &- & <3.20&- &<4.52 &4.49$\pm$1.45 & -&<63.72 &\cite{Faggi2019}\\
			2018 Jul 31 &1.16&0.59 &-  &2503$\pm$385 &- &6.16$\pm$0.12$^{(a)}$ &- &<1.80$^{(a)}$ &6.05$\pm$0.77 &- &<16.19$^{(a)}$ &\cite{Roth2020}\\
			\multicolumn{11}{c}{}\\
		    2018 Sep 07 &1.01&0.39 &3036$\pm$357 &3206$\pm$112 & -& -&- &- &10.60$\pm$1.10 &- &- &\cite{Roth2020}\\
			2018 Sep 09 &1.01&0.39 & -&2623$\pm$586 & &4.30$\pm$0.32 & &<0.62 &8.30$\pm$1.38 & &<12.59 &\cite{Faggi2019}\\
			2018 Sep 09 &1.01&0.39 &2360$\pm$33 & &4.39$\pm$0.07 & -&1.67$\pm$0.06 & -& -& 5.66$\pm$0.38& -&This Work\\
			\multicolumn{11}{c}{}\\
			2018 Oct 07 &1.07&0.49 & -&2583$\pm$864 &- &- &- & -& 4.55$\pm$1.44&- &- &\cite{Faggi2019}\\
			2018 Oct 08 &1.08&0.49 &834$\pm$25&- &1.68$\pm$0.05 &- &0.40$\pm$0.06 & -&- &1.07$\pm$0.35 & &This Work\\
			2018 Oct 10 &1.10&0.51 & -&2028$\pm$255 &- & -& -&- &2.92$\pm$0.39 &- &- &\cite{Roth2020}\\
			\hline
			\hline
	\end{tabular}}
	\tablefoot{ $^{(a)}$ From \cite{Roth2018} measured on July 29, 2018. Upper limits are 3$\sigma$ for both  \cite{Roth2020} and \cite{Faggi2019} results.}
\end{table*}

The comparison with abundances of parent molecules derived from IR data \citep{Faggi2019,Roth2020} allows us to investigate the origin of the radicals observed in 21P atmosphere. Using high-resolution infrared spectra obtained in 1998, \cite{Weaver1999} reported upper limits for different species relative to H$_2$O such as C$_2$H$_6$ (2-3\%), HCN(0.3-0.4\%) and C$_2$H$_2$(0.5-0.8\%) assuming that all species are parent molecules.
C$_2$H$_2$ has been found depleted with respect to HCN by a factor five compared to other comets like Hyakutake and Hale–Bopp, a result that has been confirmed at this apparition by \cite{Faggi2019}.

We derived a Q(C$_2$)=1.26$\times$10$^{25}$ molec/s which is consistent with the upper limit of Q(C$_2$H$_2$)<4.52$\times$10$^{25}$ molec/s reported by \cite{Faggi2019} and <1.80$\times$10$^{25}$ molec/s reported by \cite{Roth2018} at 1.18 au from the Sun. This agreement indicates that C$_2$ could  be a daughter species of C$_2$H$_2$. 
C$_2$ also has the possibility to come from C$_2$H$_6$ and HC$_2$N \citep{Helbert2005,Weiler2012,Holscher2015PhD} or released from organic-rich grains \citep{Combi1997}, but a detailed chemical model of the coma would be needed to go in more details.
We also found a good match between Q(CN)=4.40$\times$10$^{25}$ and Q(HCN)=4.30$\times$10$^{25}$ molec/s \citep{Faggi2019} at 1.01 au, showing that HCN could be  the main parent molecule of CN in 21P. This result is known for several comets using different methods, including comparison between HCN and CN production rates \citep{Rauer2003,Opitom2015}, coma morphologies \citep{Woodney2002}, and also carbon and nitrogen isotopic ratios in both species \citep{Manfroid2009,BockeleMorvan2015}. We should note however that in some cases both abundances do not agree and that other sources, for instance extended sources, have been claimed for the CN origin \citep{Fray2005}.

Some molecules such as C$_4$H$_2$, CH$_2$C$_2$H$_2$, and CH$_3$C$_2$H are proposed to be the parent molecules of C$_3$ \citep{Helbert2005,Mumma2011,Holscher2015PhD}, but these complex species were not observed at infrared or at radio wavelengths. NH and NH$_2$ were found to be depleted in 21P in the previous apparitions \citep{AHearn1995,Fink2009}. New infrared observations in 2018 show very low NH$_3$ in 21P, with an upper limit ratio of Q(NH$_3$)/Q(H$_2$O) < 0.6\% \citep{Faggi2019}. In this work, we derive  Q(NH)/Q(OH)=0.2\% which is consistent with Q(NH$_3$)/Q(H$_2$O).

\section{Optical high-resolution spectrum}
\label{sec_ratios}

\subsection{Water production rate}
\label{uves-water-production-rate}

The UVES spectrum offered the possibility of computing independently the water production rate at the time of observation. We first measured the overall flux for the OH (0,0) band near 309~nm, integrated over the whole slit. We found 1.47$\times$10$^{-13}$~erg s$^{-1}$ cm$^{-2}$ arcsec$^{-2}$.
The fluorescence efficiency computed for this band and the heliocentric distance and velocity at the time of observation was 2.62$\times$10$^{-4}$~s$^{-1}$ (or 1.71$\times$10$^{-15}$~erg s$^{-1}$ molecule$^{-1}$
if scaled to 1~au; see details on the fluorescence model in \cite{Rousselot2019}).
 From these values and a Monte-Carlo simulation of the water molecules creating OH radicals in the inner coma (model based on equations given by \cite{Combi1980}) it is possible to compute the corresponding
water production rate for the number of OH radicals observed in the slit (0.44$\times$9.5~arcsec) centered on the nucleus. Using the parameters of H$_2$O radial velocity, OH and H$_2$O lifetimes given in \cite{Cochran1993} and assuming that 91.8\% of water molecules dissociate to OH \citep{crovisier1989}, we found Q(H$_2$O)=1.7$\times$10$^{28}$ molec/s. This result is in excellent agreement with the water production rates computed from TRAPPIST observations in the same period (see Figure \ref{fig:rate_H2O}). It must, nevertheless, be pointed out that it depends of the different parameters and can change a bit with them, especially with the water lifetime.

\subsection{The $^{12}$C/$^{13}$C and $^{14}$N/$^{15}$N isotopic ratios}

The study of the isotopic ratios in comets has attracted considerable attention as it contains information about the conditions which prevailed at the time of formation of these objects in the early Solar System \citep{Jehin2009,Hyodo2013}. The carbon $^{12}$C/$^{13}$C ratio has been determined for several comets from the analysis of the C$_2$ Swan band and CN B-X system in the optical \citep[and references therein]{Manfroid2009,BockeleMorvan2015}. Some in situ measurements have also been obtained in 
comet 67P by the ROSINA mass spectrometer on-board the Rosetta spacecraft for C$_2$H$_4$, C$_2$H$_5$, CO \citep{Rubin2017} and CO$_2$ molecules \citep{Hassig2017}. All derived values are compatible with the terretrial ratio of 89, except for CO that could possibly be slightly enriched in $^{13}$C. The nitrogen $^{14}$N/$^{15}$N isotopic ratio was measured for the first time from high resolution spectra of the CN violet band in comets C/2000 WM1 (LINEAR) and C/1995 O1 (Hale-Bopp) and found to be enriched by a factor of two in $^{15}$N with respect to the Earth value \citep{Arpigny2003}. The same ratio was found later from sub-millimeter observations of HCN in comet 17P/Holmes  during its outburst and archival data of C/1995 O1 (Hale-Bopp) \citep{BockeleMorvan2008}.
It has also been recently possible to measure the $^{14}$N/$^{15}$N ratio in ammonia via the NH$_2$ radical \citep{Rousselot2014}. The values obtained are similar to the one found in HCN and CN, which was confirmed by subsequent works \citep{Shinnaka2014,Rousselot2015,Shinnaka2016a,Shinnaka2016b,Bin2018}.
Recent measurements performed by the ROSINA mass spectrometer in comet
67P provided a ratio $^{14}$N/$^{15}$N=118$\pm$25 for NH$_3$ 
and 130$\pm$30 for N$_2$ molecules \citep{Altwegg2019}.

\begin{figure*}[h!]
     \vspace{-2cm}
	 \includegraphics[scale=0.65,angle=-90]{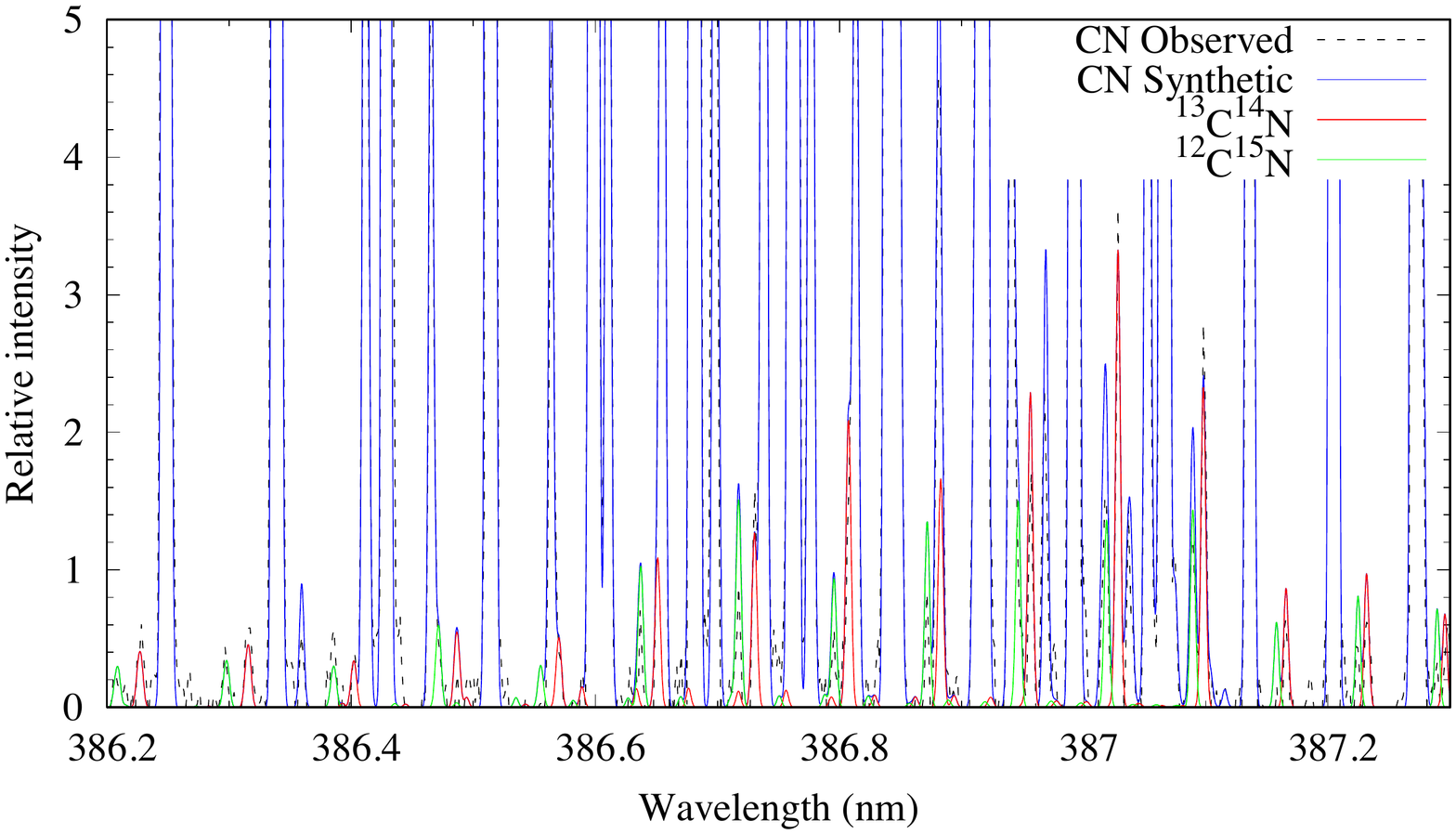}
		\vspace{-1.5cm}
	\centering \caption{The observed and synthetic CN spectra of the R branch of the B-X (0, 0) violet band in comet 21P.} 
	\label{21p_CN}

\end{figure*}

We used the $^{12}$C$^{14}$N B-X (0,0) band to estimate the $^{12}$C/$^{13}$C and $^{14}$N/$^{15}$N isotopic ratios of 21P. We used a CN fluorescence model to create synthetic spectra  of $^{13}$C$^{14}$N, $^{12}$C$^{15}$N, and $^{12}$C$^{14}$N. More details of the model are given in \cite{Manfroid2009}. Figure \ref{21p_CN} shows the observed CN spectrum compared to the synthetic one made under the same observational conditions. The ratios found for $^{12}$C/$^{13}$C and $^{14}$N/$^{15}$N are 100$\pm$10 and 145$\pm$10, respectively. These values are consistent with those of about 20 comets with different dynamical origins, 91.0$\pm$3.6 and 147.8$\pm$5.7 for $^{12}$C/$^{13}$C and $^{14}$N/$^{15}$N respectively \citep{Manfroid2009,Morvan2015}.

\subsection{NH$_2$ and NH$_3$ ortho-para ratios}

\begin{figure}[h!]
\centering \includegraphics[scale=0.36,angle=-90]{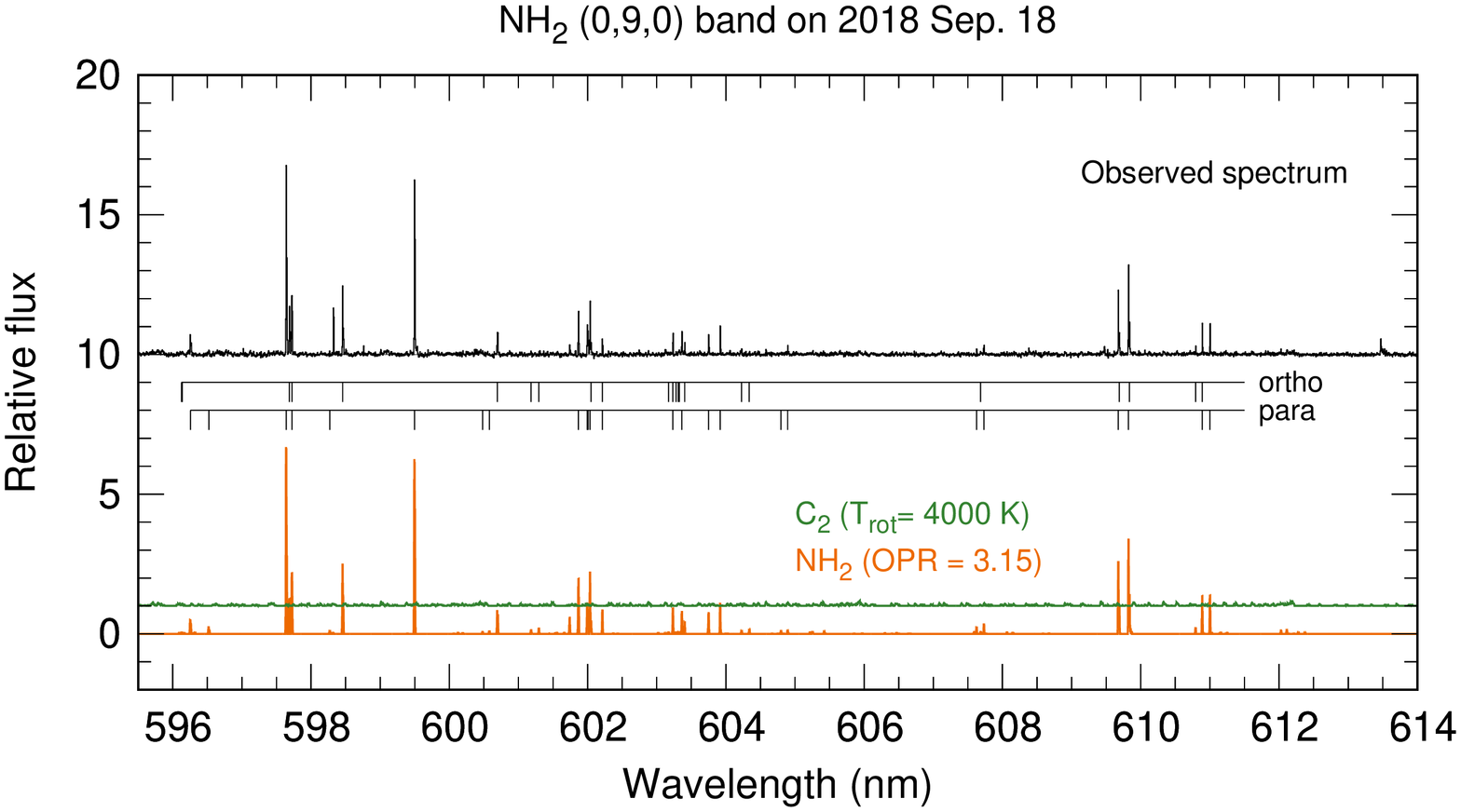}
\centering \includegraphics[scale=0.36,angle=-90]{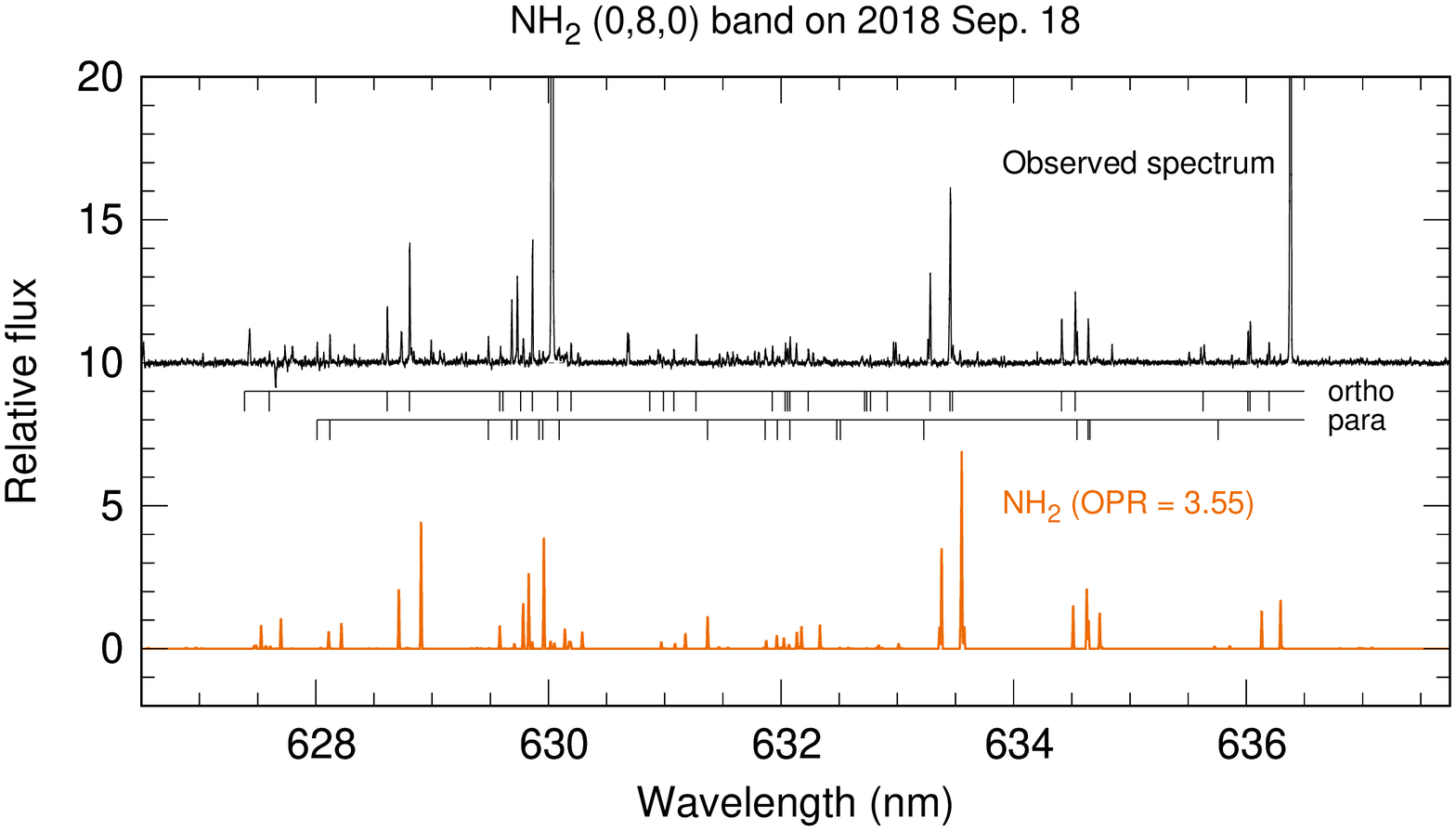}
\centering \includegraphics[scale=0.36,angle=-90]{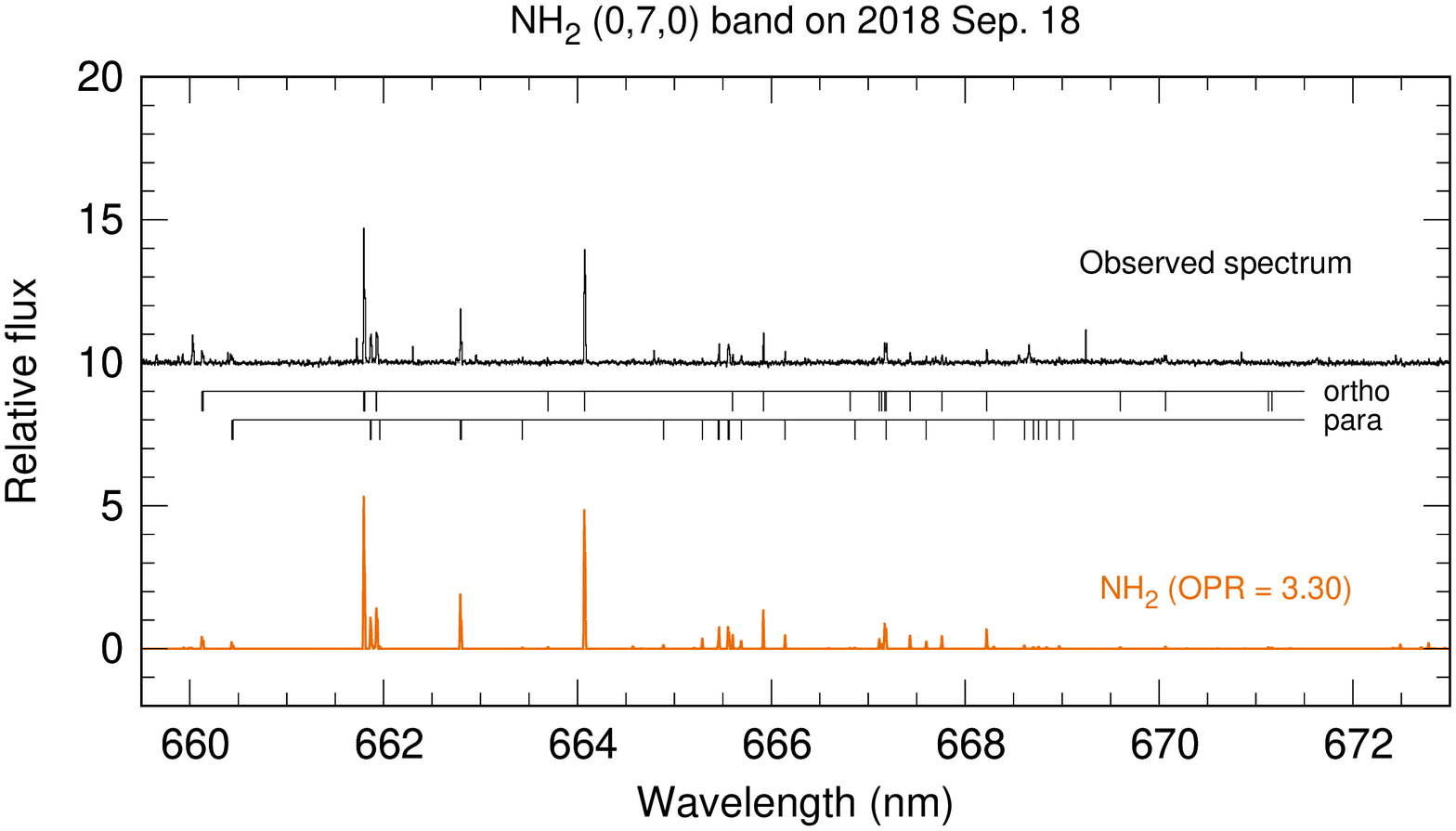}
\centering \caption{Comparison between the observed and modeled spectra of the NH$_2$(0,9,0), (0,8,0), and (0,7,0) bands. The modeled spectrum of C$_{2}$ is also plotted on the NH$_{2}$ (0,9,0) band panel, but due to the depleted nature of 21P, the C$_{2}$ lines are not affecting the NH$_{2}$ spectrum. The ortho- and para-lines of NH$_{2}$ are labeled in these modeled spectra. The two strong emission lines at 6300 \AA~ and 6364 \AA~  are the forbidden oxygen lines in the NH$_{2}$ (0,8,0) band panel. We note that the intensity ratio among bands is not correct because we scaled intensity for each plot independently.}
\label{fig:opr_nh2}
\end{figure}

\begin{figure}[h!]
	\centering	\includegraphics[width=0.72\columnwidth,angle=-90]{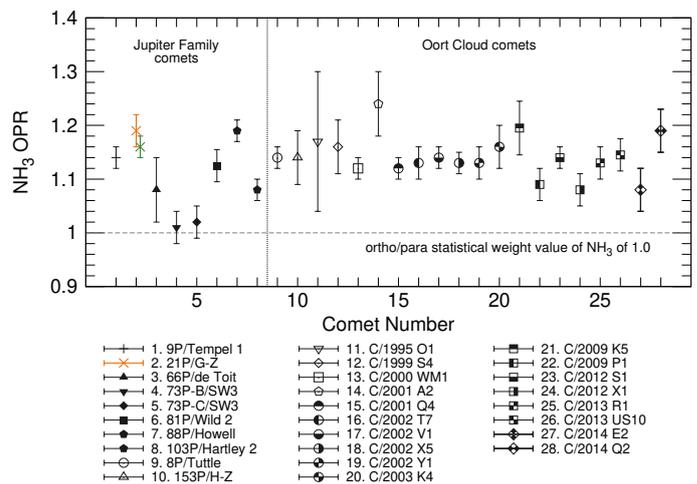}
	\caption{Summary of NH$_{3}$ OPRs in comets. The orange and green cross symbols symbols are the NH$_3$ OPR of 21P by VLT/UVES (this work) and by Subaru/HDS (\citealt{Shinnaka2020}), respectively.}
	\label{fig:OPR_NH3_comets}
\end{figure}

\begin{figure}[h!]
	\centering	\includegraphics[width=0.72\columnwidth,angle=-90]{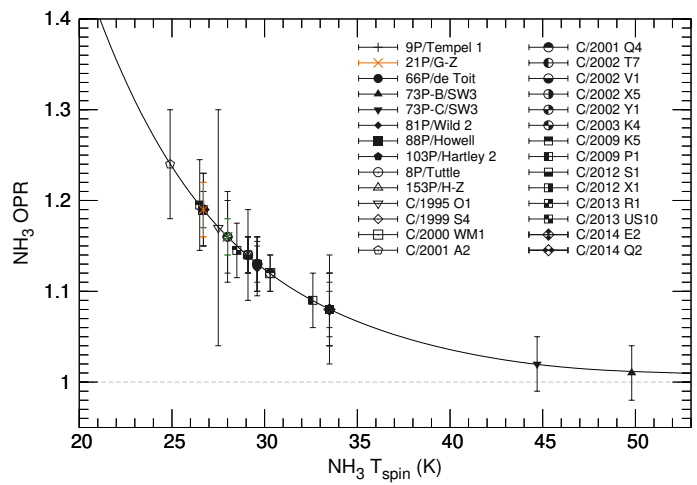}
	\caption{Summary of the NH$_{3}$ $T_{\rm spin}$ in 28 comets of various origins. The orange and green cross symbols are the NH$_3$ $T_{\rm spin}$ of 21P by VLT/UVES (this work) and by Subaru/HDS (\citealt{Shinnaka2020}), respectively.} 
	\label{fig:Tspin_NH3_comets}
\end{figure}

We measured the ortho-to-para abundance ratio (OPR) of NH$_{2}$ from the three rovibronic emissions bands (0,7,0), (0,8,0) and (0,9,0), see Figure \ref{fig:opr_nh2}, following the method described in \cite{Shinnaka2011}. The derived OPRs of NH$_{2}$ and of its parent molecule NH$_{3}$, are listed for each band in Table \ref{tab_opr} and have average values of 3.38 $\pm$ 0.06 and 1.19 $\pm$ 0.03, respectively. The latter is in very good agreement with the Subaru/HDS determination (NH$_3$ OPR = 1.16 $\pm$ 0.02; \citep{Shinnaka2020}). 
A nuclear spin temperatures ($T_{\rm spin}$) for ammonia of 27 $\pm$ 1 K was derived.
The 21P value is consistent with typical values measured in comets (see Figure \ref{fig:OPR_NH3_comets}). 21P cannot then be distinguished from other comets based on its NH$_2$ OPR (see Figure \ref{fig:Tspin_NH3_comets}), a possible cosmogonic indicator linked to the formation temperature of the molecule.

\begin{table}[h!]
\caption{Derived NH$_{2}$ and NH$_{3}$ OPRs in comet 21P}             
\label{tab_opr}      
\centering                          
\begin{tabular}{l c c c}        
\hline\hline                 
NH$_{2}$ band & NH$_{2}$ OPR & NH$_{3}$ OPR & $T_{\rm spin}$(K) \\    
\hline                        
   (0,7,0) & 3.30$\pm$0.13 & 1.15$\pm$0.07 & 28$^{+5}/_{-3}$ \\      
   (0,8,0) & 3.55$\pm$0.08 & 1.28$\pm$0.04 & 23$^{+2}/_{-1}$ \\ 
   (0,9,0) & 3.15$\pm$0.10 & 1.08$\pm$0.05 & 34$^{+8}/_{-4}$ \\
   Average & 3.38$\pm$0.06 & 1.19$\pm$0.03 & 27$\pm$1 \\
\hline                                   
\end{tabular}
\end{table}

We would like to point out that recent laboratory experiments demonstrate that the OPR of water does not keep the memory of its formation temperature \citep{Hama2011, Hama2016, Hama_Watanabe2013}. It is likely that this is also the case for ammonia. 
The OPRs of cometary volatiles might have been modified by an ortho-para conversion process in the inner coma or other catalyst activities of dust crust surfaces of the nucleus rather than reflected by a formation temperature in the solar nebula 4.6 Gy ago. OPRs might be a diagnostic of the physico-chemical conditions in the inner-most coma or beneath the surface.

\section{Dynamical evolution}
\label{sec_dynamical}
\begin{figure}[h!]
	\centering	\includegraphics[width=\hsize]{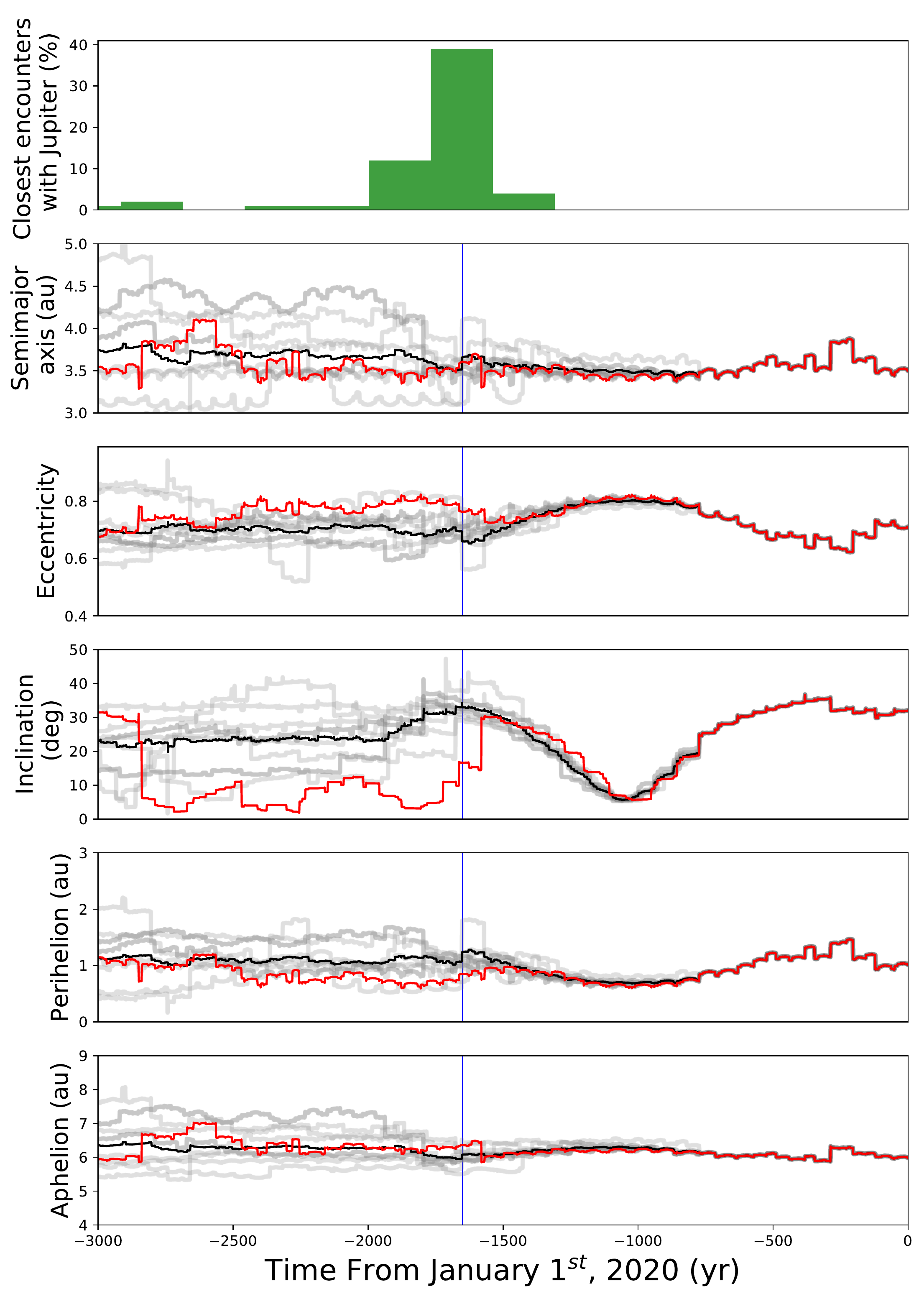}
	\caption{Orbital evolution of 21P and its 200 clones for 3000~yr backward in time from January 1, 2020. From top to the bottom: the closest approaches with Jupiter, semi-major axis, eccentricity, inclination, perihelion, and aphelion distance. In all panels, the gray lines correspond to the evolution of each clone, the black line is the mean values of the clones, and the red line is that of the nominal comet 21P. The blue-vertical line corresponds to the time of the closest encounter with Jupiter. The initial orbital elements were taken from the JPL Small-Body Data Browser (orbital solution JPL K182/3).}
	\label{dynamic}
\end{figure}

In this section we analyse the dynamical evolution of the comet within the last 10$^{5}$~yr. JFCs are highly chaotic objects, 
whose dynamic evolution must be studied in terms of statistics \citep{levison1994}. With this in mind we analysed the evolution of the original object, i.e., comet 21P, by considering the nominal values of its orbital parameters as they are defined in JPL-HORIZONS (orbital solution JPL K182/3). In the analysis, 200 clones 
were generated following the covariance matrix of its orbital parameters\footnote{Both sets of the orbital parameters and the covariance matrix of the orbit for 21P are published together in the NASA/JPL small-body browser: \url{https://ssd.jpl.nasa.gov/sbdb.cgi?sstr=21P;old=0;orb=0;cov=1;log=0;cad=0\#elem}} . We performed the integrations with 
the numerical package MERCURY \citep{chambers1999}, using the integration algorithm Bulirsch-Stoer with a time-step of 8~d and we included the Sun, all planets and Pluto in the simulation. In addition, we also included non-gravitational forces. The results of the simulations are displayed in Figure~\ref{dynamic}.  

We find that the orbits of all the clones in the simulation are very compact for a period of $\sim$1650~yr. After that period, the orbits started to scatter, which was provoked by a close encounter with Jupiter, at mean distance of 0.1~au. 
Due to the chaotic nature of JFCs, a comparison of results from different authors who applied different methods (e.g., integration algorithm used, the number of clones, how their clones were generated etc.) is difficult to perform, and any superficial comparison might yield wrong conclusions. Only one analysis identical to that performed here has been carried out: for the comet 66P/du Toit by \cite{bin2019}. 
In that study it was found that the comet belong to the Jupiter Family for at least $\sim$60$\times10^{3}$~yr, and the stable nature of its orbit was evident. 
This result hints that 21P is likely a young member of the Jupiter Family, which has crossed its perihelion $\sim$230 times with similar distances of q$\sim$1.013~au. The youth of 21P could explain its unusual composition. However, the lack of a statistic sample prevents us from robustly confirming this hypothesis.

\section{Discussion}
\label{discussion}

\begin{figure}[h!]
	\includegraphics[width=0.72\columnwidth,angle=-90]{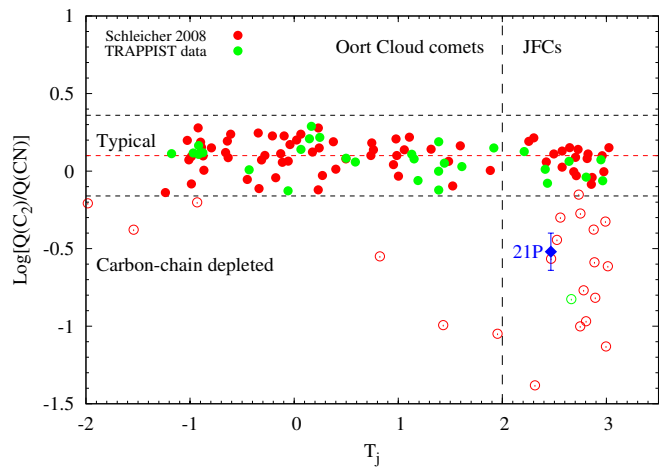}
	\vspace{-0.5cm}
	\caption{The logarithm of C$_2$-to-CN ratio of 110 comets as a function of the Tisserand invariant parameter with respect to Jupiter (T$_J$). Filled symbol present typical comets while the opened symbol present the carbon-chain depleted comets. Our measurement of comet 21P is represented by a blue diamond. The vertical dashed line at T$_j$=2 separates the families of JFCs and Oort cloud comets.}
		\label{fig:C2CN}
\end{figure}

As mentioned above, C$_2$ and  C$_3$ had been found depleted as compared to CN in 21P already more than 50 years ago \citep{Mianes1960,Herbig1976,Schleicher1987}. In \cite{AHearn1995} data set, 21P was classified as the prototype of the group depleted in carbon-chain molecules.  Figure \ref{fig:C2CN} shows our C$_2$-to-CN ratio compared to 120 comets (90 comets from \cite{Schleicher2008} and 30 comets from TRAPPIST database \citep{Opitom2016thesis}) as a function of the Tisserand invariant parameter with respect to Jupiter (T$_J$). About 30\% of the comets analyzed were found to be depleted in carbon-chain elements by varying amounts including different dynamical types of comets, two thirds are JFCs and one third are LPCs \citep{AHearn1995,Schleicher2008,Fink2009,Cochran2012}. 21P was found to be also depleted in NH with respect to OH \citep{AHearn1995}, this result was confirmed by its depletion in NH$_2$ using spectro-photometric observations by \cite{Konno1989} in 1985 apparition and later by \cite{Fink1996} in 1998 passage. This depletion in both NH and NH$_2$ indicates that 21P is likely depleted in the parent molecule NH$_3$ which was recently confirmed by \citep{Faggi2019}. 21P is not the unique case of a comet depleted in both carbon chain and ammonia daughter species. A few others have been found, but with a lesser degree of depletion like 43P/Wolf–Harrington and the split comet 73P/Schwassmann–Wachmann 3 \citep{AHearn1995,Fink2009,Cochran2012}. This indicates that there might be a small group of similar comets that formed under similar conditions and different from other comets. But according to taxonomy studies, no clear grouping associated with NH abundance has been identified.  

Our long monitoring of the abundance ratios combined with previous studies \citep{Schleicher2018,Combi2011} is ruling out the evolutionary scenario (peculiar composition due to repeated passages to perihelion). Indeed our observations show remarkably constant abundance ratios of the different species, especially the depleted C$_2$ and NH, before and after perihelion, and over months. 
The fact that these ratios are still the same after five orbits, is in favor of a pristine composition rather than compositional changes due to repeated passages of the comet at perihelion.

It was argued that this peculiar composition might be linked to a higher formation temperature, closer to the sun \citep{Schleicher1987}, or in a local disk around Jupiter as it was proposed for comet 73P \citep{Shinnaka2011}. We obtained high resolution, high SNR optical spectra in order to investigate the C and the N isotopic ratios, as well as the NH2 OPR. 21P appears to have a normal $^{14}$N/$^{15}$N ratio and a normal NH$_3$ OPR, similar to other comets. This is in contrast with comet 73P which has both peculiar $^{14}$N/$^{15}$N and OPR (see Figure \ref{fig:Tspin_NH3_comets} and \cite{Shinnaka2011}). It does not seem then that both comets are related and the peculiar composition of 21P still needs to be explained. These peculiarities are clearly linked to the ice composition of the nucleus, as the IR studies of the mother molecules are also showing the same kind of depletion, with an obvious link to the daughter species. 

\section{Summary and conclusion}
\label{sec_conclusion}

We performed an extensive monitoring of comet 21P on both sides of perihelion with TRAPPIST. The gas species production rates as well as the dust proxy, A(0)f$\rho$ parameter, were computed until the detection limit. We derived the water production rates for this apparition and we compared it, as well as the various abundance ratios, to previous passages. Using a sublimation model for the nucleus and the water production rates, we constrained the active area of the nucleus surface using slow-rotator approach. An accurate determination of the 21P nucleus parameters is needed to better constrain the active area fraction.
Comet 21P shows an asymmetric activity with respect to perihelion which might be due to  the nucleus shape, the spin axis orientation, and the distribution of activity on the comet's surface. The maximum of the gas and dust activity was about 24 days before perihelion similar to the previous apparitions. According to the molecular abundance relative to CN and OH, we confirm that 21P is depleted in C$_2$, C$_3$ and NH with respect to CN and to OH. A very good agreement between the abundance of the potential mother molecules measured in the IR (HCN, C$_2$H$_2$ and NH$_3$) and the daughter species from our optical observations has been found.
We obtained a high resolution UVES spectrum of 21P a week after perihelion and  we derived $^{12}$C/$^{13}$C and $^{14}$N/$^{15}$N isotopic ratios of 100$\pm$10 and 145$\pm$10 from the CN R-branch of the B-X (0, 0) violet band. The ammonia OPR was found equal to 1.19$\pm$0.03  corresponding to spin temperature of 27$\pm$1 K. All these values are in agreement with those found for several comets of different dynamical types and origins and do not show any peculiarity that could be related to the low carbon chain species and ammonia abundances. Our observations are favouring a pristine origin for this composition, rather than heterogeneity or evolutionary scenarios of the surface composition.

\subsection*{Acknowledgments}

{\small The research leading to these results has received funding from the ARC grant for Concerted Research Actions, financed by the Wallonia-Brussels Federation. TRAPPIST-South is a project funded by the Belgian Fonds (National) de la Recherche Scientifique (F.R.S.-FNRS) under grant FRFC 2.5.594.09.F. TRAPPIST-North is a project funded by the University of Liège, and performed in collaboration with Cadi Ayyad University of Marrakesh. E. Jehin and D. Hutsemékers are FNRS Senior Research Associates. J. Manfroid is Honorary Research Director of the FNRS. We thanks NASA, David Schleicher and the Lowell Observatory for the loan of a set of HB comet filters. UVES observations made with ESO Telescopes at the La Silla Paranal Observatory under program DDT proposal 2101.C-5051. 
}

\bibliographystyle{aa}
\bibliography{Moulane_21P.bib}
\onecolumn
\begin{appendix}

\section{Observational circumstances and production rates of comet 21P with TRAPPIST telescopes.}

\begin{table*}[h!]
	\begin{center}
		\caption{Observational circumstances of comet 21P with TRAPPIST telescopes.}
			\label{circ}
		\resizebox{0.85\textwidth}{!}{%
			\begin{tabular}{lccccccccccccccc}
				\hline
				\hline
				UT Date & $r_h$ & $\Delta$ & $\Delta$T & PA  & \multicolumn{5}{c}{Gases filters }   &  \multicolumn{5}{c}{Dust filters} & Telescope \\
				&(au) & (au) & (Days) & ($^\circ$) & OH &NH & CN & C$_2$ & C$_3$ & BC & RC & GC & Rc &  Ic & TN/TS \\
				\hline
				2018 Jun 09 &1.61 &1.07 &-93.20 &38.01 &  &   &   &   &   &   &   &  &6  &1 &TN\\
				2018 Jun 19 &1.52 &0.94 &-83.25 &40.98 &1 &   &1  &1  &1  &1  &1  &  &1  &1 &TN\\
				2018 Jun 22 &1.49 &0.90 &-80.16 &41.98 &1 &1  &1  &1  &1  &1  &1  &  &5  &1 &TN\\
				2018 Jun 28 &1.44 &0.88 &-74.20 &44.13 &1 &1  &2  &1  &1  &1  &1  &  &6  &2 &TN\\
				2018 Jul 09 &1.34 &0.78 &-63.30 &48.62 &  &   &2  &1  &   &   &1  &  &3  &1 &TN\\
				2018 Jul 26 &1.21 &0.65 &-46.25 &56.96 &1 &1  &2  &1  &1  &1  &1  &  &3  &1 &TN\\
				2018 Jul 30 &1.18 &0.62 &-42.25 &59.13 &1 &   &2  &2  &   &   &   &  &4  &  &TN\\
				2018 Aug 17 &1.07 &0.48 &-24.25 &69.44 &1 &1  &2  &1  &1  &1  &1  &  &6  &1 &TN\\
				2018 Aug 18 &1.07 &0.49 &-23.08 &69.99 &1 &   &1  &1  &1  &   &   &  &5  &1 &TN\\
				2018 Aug 23 &1.04 &0.45 &-18.13 &72.65 &1 &1  &1  &1  &1  &1  &1  &  &5  &1 &TN\\
				2018 Aug 29 &1.03 &0.42 &-12.11 &75.38 &1 &1  &1  &1  &1  &1  &1  &  &5  &1 &TN\\
				2018 Sep 05 &1.01 &0.39 &-05.10 &77.67 &1 &   &3  &1  &1  &1  &   &  &5  &1 &TN\\
				2018 Sep 09 &1.01 &0.39 &-01.10 &78.01 &1 &1  &2  &1  &1  &1  &1  &1 &8  &4 &TN\\
				2018 Sep 15 &1.01 &0.39 &+05.09 &77.40 &2 &1  &1  &1  &1  &1  &1  &  &5  &1 &TS\\
				2018 Sep 17 &1.01 &0.40 &+06.90 &76.83 &1 &1  &3  &1  &1  &1  &1  &1 &11 &2 &TN\\
				2018 Sep 17 &1.01 &0.40 &+06.90 &76.83 &1 &1  &1  &1  &1  &1  &1  &1 &5  &3 &TS\\
				2018 Sep 18 &1.01 &0.40 &+08.13 &76.53 &  &   &1  &1  &1  &1  &1  &1 &5  &3 &TS\\
				2018 Sep 20 &1.02 &0.40 &+10.12 &75.84 &1 &1  &1  &1  &1  &1  &1  &1 &5  &3 &TS\\
				2018 Sep 21 &1.02 &0.41 &+11.12 &75.45 &1 &   &1  &1  &   &1  &1  &  &3  &1 &TN\\
				2018 Sep 21 &1.02 &0.41 &+11.12 &75.45 &1 &1  &1  &1  &1  &1  &1  &1 &5  &3 &TS\\
				2018 Sep 22 &1.02 &0.41 &+12.12 &75.04 &1 &1  &1  &1  &1  &1  &1  &1 &5  &3 &TS\\
				2018 Sep 23 &1.02 &0.41 &+12.95 &74.61 &  &   &   &   &   &   &   &  &3  &1 &TN\\
				2018 Sep 25 &1.03 &0.42 &+15.10 &73.63 &1 &1  &2  &2  &1  &1  &1  &1 &5  &3 &TS\\
				2018 Sep 29 &1.04 &0.44 &+18.90 &71.61 &2 &2  &2  &2  &2  &2  &2  &  &6  &2 &TN\\
				2018 Oct 01 &1.05 &0.45 &+21.10 &70.49 &2 &1  &2  &2  &1  &1  &1  &1 &5  &3 &TS\\
				2018 Oct 02 &1.06 &0.46 &+22.10 &69.92 &1 &1  &1  &1  &1  &1  &   &  &   &  &TS\\
				2018 Oct 04 &1.07 &0.47 &+23.90 &68.69 &1 &1  &2  &2  &2  &2  &2  &  &6  &2 &TN\\
				2018 Oct 07 &1.08 &0.49 &+27.13 &66.92 &  &   &1  &1  &   &   &   &  &1  &  &TS\\
				2018 Oct 08 &1.08 &0.49 &+27.90 &66.37 &1 &1  &1  &1  &1  &1  &1  &  &11 &1 &TN\\
				2018 Oct 12 &1.10 &0.51 &+31.90 &64.55 &1 &1  &1  &2  &1  &1  &2  &  &10 &2 &TN\\
				2018 Oct 14 &1.12 &0.53 &+34.10 &62.73 &1 &1  &1  &1  &1  &1  &1  &1 &5  &1 &TS\\
				2018 Oct 15 &1.12 &0.54 &+35.10 &62.12 &  &   &   &   &   &1  &1  &1 &5  &  &TS\\
				2018 Oct 18 &1.14 &0.56 &+38.11 &60.33 &1 &1  &1  &1  &1  &1  &1  &  &3  &2 &TS\\
				2018 Oct 18 &1.14 &0.56 &+37.90 &60.33 &  &   &1  &1  &   &   &   &  &1  &  &TN\\
				2018 Oct 21 &1.16 &0.58 &+40.95 &58.55 &  &   &1  &1  &   &   &   &  &3  &  &TN\\
				2018 Oct 22 &1.17 &0.59 &+42.06 &57.97 &1 &1  &1  &1  &1  &1  &1  &  &5  &2 &TS\\
				2018 Oct 25 &1.19 &0.61 &+45.10 &56.25 &  &   &   &   &   &   &   &  &1  &1 &TS\\
				2018 Oct 28 &1.21 &0.63 &+48.09 &54.55 &  &   &1  &1  &   &1  &1  &1 &1  &1 &TS\\
				2018 Nov 01 &1.24 &0.66 &+52.04 &52.37 &1 &1  &1  &1  &1  &1  &1  &1 &4  &1 &TS\\
				2018 Nov 05 &1.27 &0.68 &+56.05 &50.80 &1 &   &1  &1  &   &1  &1  &1 &4  &1 &TS\\
				2018 Nov 09 &1.31 &0.71 &+59.98 &48.27 &1 &   &1  &1  &   &1  &1  &1 &4  &1 &TS\\
				2018 Nov 14 &1.35 &0.74 &+65.10 &45.84 &1 &   &1  &1  &   &1  &1  &1 &5  &1 &TS\\
				2018 Nov 30 &1.49 &0.85 &+81.08 &38.84 &  &   &1  &1  &   &1  &1  &  &3  &1 &TS\\
				2018 Dec 07 &1.55 &0.90 &+88.12 &36.12 &  &   &1  &   &   &1  &   &  &3  &1 &TS\\
				2018 Dec 10 &1.58 &0.92 &+91.10 &35.01 &  &   &1  &   &   &1  &   &  &3  &1 &TS\\
				2018 Dec 19 &1.66 &0.98 &+100.05&32.04 &  &   &1  &   &   &1  &1  &  &4  &1 &TS\\
				2018 Dec 29 &1.75 &1.06 &+110.12&29.28 &  &   &1  &   &   &   &   &  &1  &1 &TS\\
				2019 Jan 14 &1.90 &1.20 &+126.13&26.20 &  &   &   &   &   &   &   &  &2  &  &TS\\
				2019 Jan 15 &1.91 &1.21 &+127.10&26.08 &  &   &   &   &   &   &   &  &3  &1 &TS\\
				2019 Jan 29 &2.04 &1.37 &+141.12&24.74 &  &   &   &   &   &   &   &  &4  &1 &TS\\
				2019 Feb 02 &2.08 &1.42 &+145.08&24.52 &  &   &   &   &   &   &   &  &4  &1 &TS\\
				2019 Feb 04 &2.10 &1.44 &+147.11&24.43 &  &   &   &   &   &   &   &  &4  &4 &TS\\
				\hline	
				\hline
		\end{tabular}}
	\end{center}
	\tablefoot{$r_h$ and $\bigtriangleup$ are respectively the heliocentric and geocentric distances, $\Delta$T is the time to perihelion in days, (-) for pre-perihelion and (+) for post-perihelion. PA is the Solar phase angle. }
\end{table*}

\begin{table*} 
	\begin{center}
		\caption{OH, NH, CN, C$_2$, and C$_3$ production rates and A($\theta$=0)f$\rho$ measurements for comet 21P with both TN and TS telescopes.}
		\label{tab:rates}
		\resizebox{\textwidth}{!}{%
			\begin{tabular}{lcccccccccccc}
				\hline
				\hline
UT Date & $r_h$  & \multicolumn{5}{c}{Production rates ($\times$10$^{24}$ molec/s) }   &  \multicolumn{5}{c}{A($\theta$=0)f$\rho$} & TN/TS\\
	    &(au)               & OH       & NH       & CN      & C$_2$       & C$_3$            & BC   & RC   & GC   & Rc   &  Ic & \\
				\hline
2018 Jun 09 &1.61  & -& - &  - & -  &  - &   -&   -&  -&236.5$\pm$14.0  &265.5$\pm$18.7 &TN\\
2018 Jun 19 &1.52 &- & - &25.20$\pm$0.52 &4.30$\pm$0.65 & - &234.8$\pm$13.6  & 351.1$\pm$17.5& - & - &326.9$\pm$15.6 &TN\\
2018 Jun 22 &1.49 & 6200$\pm$253& - &28.00$\pm$0.53  & - &1.50$\pm$0.20  & - & -& - &333.6$\pm$16.8  &367.5$\pm$18.4 &TN\\
2018 Jun 28 &1.44 & 9560$\pm$365& - &32.00$\pm$0.64  & - & - & - & 673.7$\pm$21.4 & - &411.0$\pm$13.7  &475.7$\pm$12.9 &TN\\
2018 Jul 26 &1.21 &15200$\pm$322 & 31.30$\pm$4.90&37.80$\pm$0.54  &8.40$\pm$0.66  &1.08$\pm$0.20  & 618.6$\pm$36.4 & 801.5$\pm$12.9 & - & 687.8$\pm$11.5 &758.2$\pm$15.9 &TN\\
2018 Jul 30 &1.18 &18100$\pm$278  & - &45.50$\pm$0.63  &12.60$\pm$0.65  & -  & -  &-   & - &959.8$\pm$15.8  & -&TN\\
2018 Aug 17 &1.07 &28300$\pm$558&54.10$\pm$4.59  &52.00$\pm$0.65  &17.90$\pm$0.64  &2.52$\pm$0.19  &1087.6$\pm$13.7 & 1646.1$\pm$12.7 & - &1502.7$\pm$12.1  &1609.3$\pm$12.7&TN\\
2018 Aug 18 &1.07 &25600$\pm$276 &  - &50.80$\pm$0.64  &18.00$\pm$0.62  &2.47$\pm$0.17  & -  & -  &-  &1522.1$\pm$11.6  &1648.9$\pm$10.3 &TN\\
2018 Aug 23 &1.04 &27300$\pm$304 &50.20$\pm$3.60  &55.20$\pm$0.67  &19.20$\pm$0.64  & 2.55$\pm$0.17 & 1028.9$\pm$12.5& 1570.7$\pm$11.8 & - &1463.7$\pm$11.7  &1570.2$\pm$10.0 &TN\\
2018 Aug 29 &1.03 &23000$\pm$286 &43.50$\pm$4.07  &42.50$\pm$0.62  &13.70$\pm$0.62  &1.66$\pm$0.18  & 887.6$\pm$29.6 & 1229.9$\pm$12.5 & - &1100.0$\pm$12.7  &1176.3$\pm$16.1 &TN\\
2018 Sep 05 &1.01 &20800$\pm$312 &  - &38.20$\pm$0.64  &14.00$\pm$0.63  &1.95$\pm$0.17  & 651.2$\pm$12.4 & -  & - &933.1$\pm$10.5  &984.0$\pm$14.8 &TN\\
2018 Sep 09 &1.01 &23600$\pm$328 &56.60$\pm$3.86  &43.90$\pm$0.68  &16.70$\pm$0.64  &2.26$\pm$0.17  &723.3$\pm$12.4  & 1103.1$\pm$15.7 & 815.5$\pm$12.4&1062.9$\pm$12.7  &1123.3$\pm$12.1 &TN\\
2018 Sep 17 &1.01 &17100$\pm$300 & 39.80$\pm$3.77 & 37.10$\pm$0.66 &11.60$\pm$0.64  &1.91$\pm$0.18  & 574.6$\pm$14.1 & 861.6$\pm$12.4 &642.0$\pm$10.7&799.0$\pm$15.1 &833.7$\pm$14.0 &TN\\
2018 Sep 17 &1.01 &14700$\pm$326 &38.30$\pm$3.84  &37.60$\pm$0.53  &12.30$\pm$0.63   &2.02$\pm$0.22  &571.6$\pm$19.6  &851.4$\pm$17.2 &599.9$\pm$19.2 &767.0$\pm$16.2  &832.1$\pm$14.8&TS\\
2018 Sep 18 &1.01 &- & - &30.00$\pm$0.52  &11.00$\pm$0.65  &2.07$\pm$0.28  &459.5$\pm$34.2 &872.2$\pm$22.2  &595.9$\pm$14.8 &733.3$\pm$14.8  &849.6$\pm$20.4 &TS\\
2018 Sep 21 &1.02 & -&-  & - & - & - &-  & - &  -&604.7$\pm$16.6  &652.2$\pm$14.9 &TN\\
2018 Sep 21 &1.02 &1280$\pm$212 &28.50$\pm$3.79  &33.00$\pm$0.52  &10.90$\pm$0.63  &1.72$\pm$0.22  &542.3$\pm$17.4  &808.3$\pm$12.5 &606.4$\pm$19.9 &658.3$\pm$14.9 &779.5$\pm$21.4 &TS\\
2018 Sep 22 &1.02 &13600$\pm$216 &29.40$\pm$3.84  &32.80$\pm$0.51  &10.80$\pm$0.65  &1.64$\pm$0.23  &510.9$\pm$17.4  &744.0$\pm$15.0  &569.2$\pm$12.5 &649.8$\pm$12.5  &740.3$\pm$14.7 &TS\\
2018 Sep 23 &1.02 &12800$\pm$249 &  - &28.60$\pm$0.61   & 9.48$\pm$0.62  &  - &394.5$\pm$12.5   & 596.1$\pm$12.5  & - & - &629.3$\pm$15.2 &TN\\
2018 Sep 25 &1.03 &10880$\pm$216 & 29.00$\pm$4.15 & 27.30$\pm$0.54 &9.62$\pm$0.76  &1.24$\pm$0.31  &406.8$\pm$22.4  &611.7$\pm$19.9  &460.0$\pm$20.1 &540.5$\pm$14.8  &611.6$\pm$15.3 &TS\\
2018 Sep 29 &1.04 &11300$\pm$254 &23.80$\pm$4.11  &24.50$\pm$0.58  &7.21$\pm$0.67  &0.79$\pm$0.19  &368.1$\pm$20.4  & 517.1$\pm$15.1 &  &530.7$\pm$17.8  &535.4$\pm$15.2 &TN\\
2018 Oct 01 &1.05 &10220$\pm$217 &18.40$\pm$3.68  &22.40$\pm$0.52  &7.84$\pm$0.64  &1.13$\pm$0.23  & 381.4$\pm$23.2 & 490.7$\pm$20.6 & 368.1$\pm$21.9 & 418.3$\pm$15.3 & 478.8$\pm$15.7 &TS\\
2018 Oct 02 &1.06 &9860$\pm$182 &19.50$\pm$3.27  &21.70$\pm$0.52  &6.55$\pm$0.71  &1.04$\pm$0.21  &350.8$\pm$15.3  & - & - & - & - &TS\\
2018 Oct 04 &1.07 &9460$\pm$268&  &20.20$\pm$0.55  & 5.95$\pm$0.63 &0.70$\pm$0.18  &282.6$\pm$14.0 &412.3$\pm$12.2  &  -&411.2$\pm$12.2  &437.4$\pm$15.4 &TN\\
2018 Oct 08 &1.08 &8340$\pm$248&10.70$\pm$3.53  &16.80$\pm$0.53  &4.03$\pm$0.61 &0.36$\pm$0.17  & -& 318.5$\pm$19.9 & - &350.5$\pm$12.2 &350.5$\pm$18.3 &TN\\
2018 Oct 12 &1.10 &5950$\pm$248&  -&15.80$\pm$0.52  & 4.24$\pm$0.63 & - & 199.7$\pm$12.3& 280.1$\pm$19.1& - &294.6$\pm$12.8 &296.5$\pm$20.4 &TN\\
2018 Oct 14 &1.12 &4730$\pm$247&11.00$\pm$4.16  &13.30$\pm$0.52  &3.90$\pm$0.67 & -&190.9$\pm$25.7 & - &- &234.0$\pm$17.3  &269.8$\pm$20.5 &TS\\
2018 Oct 15 &1.12 & - &  - & -  &  - & -  &166.9$\pm$18.3  &257.9$\pm$18.6 &192.1$\pm$21.2 &226.1$\pm$13.3  & - &TS\\
2018 Oct 18 &1.14 &- &  -&  -& - &-  &151.0$\pm$17.2 & - &-  &208.2$\pm$12.4  &230.8$\pm$11.7 &TS\\
2018 Oct 18 &1.14 & - & -  &10.80$\pm$0.54  & - & -  & -  & -  &  -&229.7$\pm$14.3 & - &TN\\
2018 Oct 21 &1.16 & - & -  &9.87$\pm$0.51  & 3.02$\pm$0.65 & -  &  - &  - &-  &198.2$\pm$19.4 &  -&TN\\
2018 Oct 22 &1.17 &3500$\pm$188 &  &9.04$\pm$0.50 &2.30$\pm$0.65  &0.34$\pm$0.21  &137.6$\pm$21.1  &149.0$\pm$26.1  & - &114.4$\pm$13.3  &127.9$\pm$14.4 &TS\\
2018 Oct 25 &1.19 &  -& -  &  - &  - & -  &  - &  - & - &111.5$\pm$20.0  &- &TS\\
2018 Oct 28 &1.21 & - & -  &6.00$\pm$0.55  & -&  - & - &151.2$\pm$13.8  &117.2$\pm$16.5&137.6$\pm$12.1 &144.7$\pm$14.1 &TS\\
2018 Nov 01 &1.24 &2320$\pm$234 & - &6.88$\pm$0.56  &  -& - & - &108.1$\pm$18.6  &85.7$\pm$23.6 &109.6$\pm$15.7  &123.2$\pm$12.4 &TS\\
2018 Nov 05 &1.27 &2010$\pm$191 &  - &5.62$\pm$0.52  &1.16$\pm$0.66  & -  &95.7$\pm$20.4 &120.5$\pm$21.8  &50.1$\pm$16.2&121.1$\pm$15.4  &139.0$\pm$15.0 &TS\\
2018 Nov 09 &1.31 &- & -  & - & - &  - &- &- & -&88.8$\pm$13.4 &129.3$\pm$15.6&TS\\
2018 Nov 14 &1.35 &- & -  &4.39$\pm$0.55 &1.27$\pm$0.69  &  - &85.5$\pm$13.2  &88.7$\pm$18.7  &87.2$\pm$14.6 &86.9$\pm$13.6 &91.3$\pm$12.0 &TS\\
2018 Dec 07 &1.55 &  -& -  &3.30$\pm$0.52  & -  &   -&- &  - & - &42.9$\pm$10.5  &71.7$\pm$13.2 &TS\\
2018 Dec 10 &1.58 & - & -  &3.00$\pm$0.56  & -  &  - &  -& 75.9$\pm$14.2  &  &28.1$\pm$13.9  &-&TS\\
2018 Dec 19 &1.66 & - & -  &3.30$\pm$0.54  & -  & -  & - &80.6$\pm$12.5  &  &27.2$\pm$11.5 &172.3$\pm$18.2&TS\\
2018 Dec 29 &1.75 & - & -  &-  &  - &  - &  - & -  &  -&-  &57.9$\pm$18.1 &TS\\
2019 Jan 14 &1.90 &- &-   &  -&  -& - &-  & - &  -&60.7$\pm$11.1  & -&TS\\
2019 Jan 15 &1.91 & -&  - & - & - & - &  -& - & - &57.2$\pm$11.0  &60.4$\pm$11.6&TS\\
2019 Jan 29 &2.04 &- &  - & - &  -&  -& - &  -&  -&85.5$\pm$11.4  &93.5$\pm$11.5 &TS\\
2019 Feb 02 &2.08 & -& -  &  -& - & - & - & - &-  &83.7$\pm$12.5  &91.6$\pm$11.3&TS\\
2019 Feb 04 &2.10 & -& -  &  -& - & - & - &  -& - & 71.5$\pm$22.0 &96.3$\pm$17.1&TS\\
			\hline	
				\hline
		\end{tabular}}
	\end{center}
	\tablefoot{$r_h$ is the heliocentric distances. The A(0)f$\rho$ values are printed at 10000 km form the nucleus and corrected from the phase angle effect.}
\end{table*}

\end{appendix}
\end{document}